\documentclass[journal]{IEEEtran}
\usepackage{amsmath,amsfonts}
\usepackage{array}
\usepackage[caption=false,font=normalsize,labelfont=sf,textfont=sf]{subfig}
\usepackage{textcomp}
\usepackage{stfloats}
\usepackage{url}
\usepackage{verbatim}
\usepackage{graphicx}
\usepackage{booktabs}
\allowdisplaybreaks  
\usepackage{cite}
\usepackage{tikz}
\usepackage{pgfplots}
\pgfplotsset{compat=1.17}
\usepackage{xcolor}
\usepackage{bm}
\usepackage{upgreek}
\usepackage{hyperref}
\usepackage{textcomp} 
\usepackage{multirow}
\usepackage{fix-cm}    
\usepackage[table]{xcolor}
\usepackage{calc} 
\usepackage{mathtools}
\usetikzlibrary{calc}
\usepackage{pgf-pie} 

\usepackage{siunitx} 
\begin{document}

\title{Pricing Short-Circuit Current via a Primal-Dual Formulation for Preserving Integrality Constraints}

\author{Peng Wang,~\IEEEmembership{Student Member,~IEEE}, and Luis Badesa
    \vspace*{-1cm}    
\thanks{Peng Wang and Luis Badesa are with the School of Industrial Engineering and Design (ETSIDI), Technical University of Madrid (UPM), Spain (e-mail: peng.wang@alumnos.upm.es, luis.badesa@upm.es). \textit{(Corresponding author: Luis Badesa.)}}
}

\maketitle

\begin{abstract}
Synchronous Generators (SGs) currently provide important levels of Short-Circuit Current (SCC), a critical ancillary service that ensures line protections trip during short-circuit faults. Given the ongoing replacement of SGs by power-electronics-based generation, which has a hard limit on current injection, it has become relevant to optimize the procurement of SCC services provided by remaining SGs. Pricing this service, i.e., computing dual variables of SCC constraints, is however challenging due to the integrality constraints in Unit Commitment (UC). Existing methods, e.g., dispatchable pricing and restricted pricing, attempt to address this issue but exhibit limitations in handling non-convexities, resulting in SCC prices that either fail to cover the operating costs of units or lack interpretability. To overcome these pitfalls, we adopt a primal-dual formulation of the existing SCC-constrained dispatch that preserves the binary UC for effectively computing shadow prices of SCC services. Using a modified IEEE 30-bus system, the proposed method is compared with the previously developed pricing schemes. It is demonstrated that, under the proposed pricing method, revenue-adequate and explicit service prices can be assigned without the need for make-whole payments, an advantage that cannot be achieved by other pricing approaches.
\end{abstract}

\begin{IEEEkeywords}
Ancillary services, short-circuit current, primal-dual formulation, shadow prices, unit commitment.
\end{IEEEkeywords}

\vspace{-0.2cm}

\section*{Nomenclature}
\addcontentsline{toc}{section}{Nomenclature}

\subsection*{Abbreviations and Acronyms}
\begin{IEEEdescription}
\item[IBR] Inverter-Based Resources
\item[LP] Linear Program
\item[MILP] Mixed-Integer Linear Program
\item[P-D] Primal-Dual   
\item[SCC] Short-Circuit Current 
\item[SG] Synchronous Generator     
\item[UC] Unit Commitment
\end{IEEEdescription}

     \vspace{-0.4cm}
\subsection*{Indices and Sets}
\begin{IEEEdescription}
    \item [$\Psi(g)$, $\Phi(c)$] \qquad Indices of buses with SGs and IBR
    \item[$b, \mathcal{B}$] Index, Set of buses
    \item[$c, \mathcal{C}$] Index, Set of IBR
    \item[$g,\mathcal{G}$] Index, Set of SGs   
    \item[$m, \mathcal{M}$]  Index, Set for pairs of commitment variables
    \item[$t, T$]  Index, Set of periods for system operation
\end{IEEEdescription}

     \vspace{-0.4cm}
\subsection*{Constants and Parameters}
\begin{IEEEdescription}
    \item[$\alpha_{c}$] Capacity percentage of IBR 
    \item[$\mathrm{c}_g^\mathrm{m}$]  Marginal generation cost of SG (\texteuro/MWh)
    \item[$\mathrm{c}_g^\mathrm{nl}$]  No-load cost of SG (\texteuro/h) 
    \item[$\mathrm{c}^\mathrm{st}_g$] Startup cost of SG (\texteuro/h) 
    \item[$\mathrm{I}_{b_\mathrm{lim}}$] Minimum requirement of SCC for bus $b$ (p.u.)
    \item[$\mathrm{k}_{bg}, \hspace{-1pt}\mathrm{k}_{bc}, \hspace{-1pt}\mathrm{k}_{bm}$] \qquad~ Coefficients of approximate SCC (p.u.)
    \item[$\mathrm{P}^\mathrm{D}$] Total energy demand of system (MW) 
    \item[$\mathrm{P}_g^\mathrm{min}$] Minimum stable generation of SG (MW)
    \item[$\mathrm{P}_g^\mathrm{max}$] Rated power of SG (MW)
    \item[$\mathrm{P}^{\mathrm{max}}_{c}$]  Rated power of IBR (MW) 
\end{IEEEdescription}

     \vspace{-0.4cm}
\subsection*{Primal variables} 
\begin{IEEEdescription}
    \item[$C_{g}^\mathrm{st}$] Startup cost incurred by SG (\texteuro/h)
    \item[$P_{g}$] Power output of SG  (MW)
    \item[$P_{c}$] Power output of IBR (MW)
    \item[$u_{g}$] Binary variable, commitment state of SG
    \item[$\eta_{m}$] Binary variable, product of two commitment states 
\end{IEEEdescription}

     \vspace{-0.4cm}
\subsection*{Dual variables (only shadow prices are listed)}
\begin{IEEEdescription}
    \item[$\lambda^\mathrm{E}$] Energy price (\texteuro/MWh) 
    \item[$\lambda_{b}^{\mathrm{SCC}}$]  SCC price for bus $b$ (\texteuro/p.u.)
    \item[$\lambda_{g,\mathrm{commit}}$] \quad~ Commitment price for SG (\texteuro/h) 
\end{IEEEdescription}

\vspace{-0.2cm}
\section{Introduction}
\IEEEPARstart{P}OWER grids worldwide are undergoing a transition toward renewable energy-dominated architectures in pursuit of net-zero emissions. This transition inevitably leads to a higher penetration of weather-driven Inverter-Based Resources (IBR), which in turn raises the need for new stability services to ensure operational security \cite{chaudhuri2024rebalancing,ali2025power}. Regarding short-circuit faults, protection devices can safeguard the system only when sufficient Short-Circuit Current (SCC) is detected by the relays, necessitating that the SCC at all system buses is maintained at the level required by the relays. However, IBR inherently provide very limited SCC due to their restricted over-current capability \cite{jia2017review}. At the same time, conventional Synchronous Generators (SGs), which are capable of delivering relatively high SCC \cite{tleis20197}, are gradually being phased out in favor of renewable energy sources. As a result, the system-wide SCC level is expected to decline \cite{kubis2014development}, making the provision of sufficient SCC a critical challenge that directly affects the reliability of short-circuit fault detection.

Relevant studies have been conducted to ensure reliable protection operation. In \cite{ferrari2024grid}, inverter short-circuit capacity is enhanced and control strategies are optimized to deliver sufficient SCC. Studies in \cite{yang2015optimal,tang2022optimal} show that SCC levels can be regulated through transmission line switching to modify system topology. Reference \cite{xiao2024optimization} optimizes relay parameters to adapt to load current variations and improve protection sensitivity, while \cite{wang2025adaptive} adjusts protection settings according to fault scenarios and distributed generation penetration. 

Although these methods support efficient relay operation, they neglect the SCC contribution from synchronous machines, which can help alleviate insufficient SCC levels. To fill this gap, reference \cite{chu2021short} introduces an SCC constraint considering current injections from both SGs and IBR, allowing units to optimize their operating points and maintain required SCC levels for reliable relay operation. This method leverages the steady-state SCC capability of generating units as a potential ancillary service. Existing electricity markets do not currently operate dedicated market clearing mechanisms for SCC as a standalone service, though SCC is formally recognized in legal frameworks and academic literature as an emerging non-frequency control ancillary service in the EU \cite{chu2024pricing,wang2026imperfect,EUreport}. 

Corresponding pricing frameworks for this service are proposed in \cite{chu2024pricing}, where the main challenge lies in handling the non-convexity introduced by Unit Commitment (UC) embedded in SCC constraints, as binary variables prevent the derivation of marginal prices from dual variables. Previously proposed pricing approaches for addressing binary variables when computing shadow prices include the so-called `dispatchable pricing' and `restricted pricing'. Dispatchable pricing relaxes binary commitment decisions into continuous variables, thereby enabling the derivation of dual variables for SCC constraints. Restricted pricing seeks to determine a `commitment price' representing the economic value of SCC provision, which is then bundled with the value of other services that only depend on the on/off status of a unit, such as inertia \cite{badesa2022assigning}. Another pricing method based on non-dual theory is `marginal-unit pricing'; it evaluates a unit's economic value for SCC ancillary services by alternately removing its contribution to the SCC constraint and calculating the resulting difference in system redispatch costs. Although these approaches provide important insights, each has limitations on its applicability.

The dispatchable method sacrifices the discreteness of UC decisions, an essential physical condition for the system. Consequently, the McCormick envelopes used to linearize binary variable products within SCC expressions cannot precisely capture the actual SCC level, and these bilinear terms therefore need to be eliminated. Moreover, relaxing integrality would result in prices that do not support market equilibrium if adequate uplift payments are not provided to relevant generators, in the sense that these producers may not wish to produce under such prices as they would incur losses. The restricted method has the important disadvantage of conflating the price of any service related to the commitment of a unit, such as SCC and inertia, into a single dual variable, leading to low interpretability of the economic incentives for different services. This would be an important limitation for remunerating synchronous condensers, as demonstrated in \cite{badesa2022assigning}. The marginal-unit method does not involve the calculation of dual variables, and it fails to yield explicit price signals for ancillary services and energy; consequently, payments for the unit cannot be settled, nor can the method be effectively integrated into a market framework. Therefore, this method would be only suitable for assessing how well a specific unit meets the demand for SCC services at a given node.

To overcome these limitations, this paper proposes a new SCC pricing approach based on the Primal-Dual (P-D) formulation introduced by \cite{ruiz2012pricing}. The method first relaxes all binary variables of the primal to continuous ones, then formulates its dual along with constraints that explicitly guarantee non-negative profits for dispatched units. Finally, it enforces the relaxed variables back to discrete values while solving a problem that minimizes the duality gap. In this manner, service prices that deviate minimally from those under integrality relaxation can be derived and effectively provide generators with the proper incentives to remain in the market.

It should be noted that prior works \cite{chu2024pricing} and \cite{wang2026imperfect} have already developed SCC‑constrained UC and established the general concept of SCC ancillary‑service pricing. This paper does not aim to propose new SCC‑constrained dispatch modeling. Instead, it focuses on resolving a remaining open challenge: how to compute interpretable locational SCC shadow prices while preserving binary commitment variables and guaranteeing generator revenue adequacy without make-whole payments. Specifically, the main contributions of this work are:
\begin{enumerate}
    \item To adopt, for the first time, a P-D non‑convex pricing formulation for the SCC ancillary service, which critically retains the discrete nature of UC and an accurate physical SCC representation.
    \item To establish a rigorous mathematical framework for directly deriving revenue-adequate shadow prices. It is shown that relevant generators can achieve non-negative profits under such prices, without relying on make-whole payments required by alternative approaches.
    \item To demonstrate that previously proposed pricing methods would either induce misleading SCC signals for unnecessary buses and underestimated SCC prices at critical buses, leading to insufficient SCC revenue to cover generator losses; or be completely unsuitable to provide explicit SCC prices for relevant devices such as synchronous condensers.
\end{enumerate}

The remainder of this paper is organized as follows: Section \ref{Review of Existed Short-Circuit Current Pricing Schemes} introduces the SCC constraint and reviews the state-of-the-art SCC pricing methods. Section \ref{Primal-Dual Formulation for Pricing SCC Services} presents the proposed pricing framework and its mathematical formulation via a general SCC-constrained UC. Section \ref{Case Studies} includes case studies that showcase the advantages of the proposed approach. Finally, Section \ref{Conclusion} concludes the paper and outlines future research.

\section{Review of Existing Studies for Pricing Short-Circuit Current Services}\label{Review of Existed Short-Circuit Current Pricing Schemes}
This section begins by introducing the SCC constraints. Note that only three-phase nodal short-circuit faults are considered in this study. It then reviews previously proposed SCC pricing methods, identifying their limitations in addressing non-convexities. Finally, basic economic principles and settings applied to the proposed market are clarified.

\vspace{-0.2cm}
\subsection{Representation of SCC Constraints }\label{Formulation of Short Circuit Current Constraints}
This work adopts the SCC constraints derived in \cite{chu2021short}, which incorporate the current contributions from both SGs and IBR. For a power system with SGs $g \in \mathcal{G}$ and IBR $c \in \mathcal{C}$, the SCC level at bus $b$ can be formulated as:
\begin{equation}
     I_{b_\mathrm{SC}}= \frac{\sum_{g\in\mathcal{G}}Z_{b\Psi(g)}\mathrm{I}_gu_g+\sum_{c\in\mathcal{C}}Z_{b\Phi(c)}\mathrm{I}_c\alpha_c}{Z_{bb}}  \label{eq:define_of_SCC_constraints}  
\end{equation}
which captures the discrete nature of SG-provided SCC via the commitment variable $u_g$. Furthermore, the SCC contribution also depends on the impedance $Z_{ij}$ and, in the case of IBR, on their capacity percentage $\alpha_c$. $\mathrm{I}_g$ and $\mathrm{I}_c$ denote the short-circuit injections from SGs and IBR, respectively. The improvement in SCC levels across buses from this constraint is shown in Sections~\ref{SCC Improvement at Risk Buses} and \ref{SCC Improvement at Risky Buses_multiperiod}.

The impedance matrix is obtained by inverting the corresponding admittance matrix, making $Z_{ij}$ difficult to incorporate into duality-involved optimization. A data-driven approach is thus employed to approximate the actual SCC constraints by optimizing coefficients $\mathcal{K} = \{k_{bg},k_{bc},k_{bm}\}$ \cite{chu2021short}:
\begin{subequations} \label{eq:SCC_constraints_linearlized}
\begin{align}
& I_{b_\mathrm{L}}= \sum_{g}k_{bg}u_{g}+\sum_{c}k_{bc}\alpha_{c}+\sum_{m}k_{bm}\eta_{m} \ge \mathrm{I}_{b_{\mathrm{lim}}} \label{eq:define_SCC_constraints_linearized} \\
& \eta_m=u_{g_\mathrm{1}} \cdot u_{g_\mathrm{2}},\quad\mathrm{s.t.}~\{g_\mathrm{1},g_\mathrm{2}\}=m    \label{eq:interactions_pair_SGs_1}         \\ 
& m\in\mathcal{M}=\{g_{\mathrm{1}},g_{\mathrm{2}}\mid \forall g_{\mathrm{1}}, g_{\mathrm{2}}\in\mathcal{G}\}    \label{eq:interactions_pair_SGs_2}    
    \end{align}
\end{subequations}
where \eqref{eq:define_SCC_constraints_linearized} defines the approximated SCC, which is enforced to be greater than $\mathrm{I}_{b_\mathrm{lim}}$. $\eta_m$ models the pairwise interactions between SGs to emulate nonlinear behavior, defined as~\eqref{eq:interactions_pair_SGs_1}-\eqref{eq:interactions_pair_SGs_2}. The coefficients $\mathcal{K}$ are determined via an optimization-based classification procedure, which involves enumerating system operating points and minimizing the error introduced by the approximation, as in \eqref{eq:opt_problem}. While $\alpha_c$ theoretically spans the continuous interval $[0, 1]$, restricting it to $\alpha_c \in \{0, 1\}$ would adequately capture IBR behavior without requiring daily parameter updates \cite{chu2021short}. The detailed implementation of this optimization is given below:
\begin{subequations}\label{eq:opt_problem}
\begin{align}
& \min_{\mathcal{K}} \sum_{\omega \in \Omega_2}
\left( I_{b_\mathrm{L}}^{(\omega)} - I_{b_\mathrm{SC}}^{(\omega)} \right)^2
\label{eq:opt_problem_a} \\
 &\mathrm{subject~to:} \nonumber \\
& \hspace{5mm}I_{b_\mathrm{L}}^{(\omega)} < \mathrm{I}_{b_{\mathrm{lim}}},~ \forall \, \omega \in \Omega_1
\label{eq:opt_problem_b} \\
& \hspace{5mm}I_{b_\mathrm{L}}^{(\omega)} \ge \mathrm{I}_{b_{\mathrm{lim}}},~ \forall \, \omega \in \Omega_3
\label{eq:opt_problem_c} \\
& \hspace{5mm}\Omega = \Omega_1 \cup \Omega_2 \cup \Omega_3
\label{eq:opt_problem_d} \\
& \hspace{5mm}\Omega_1 =
\left\{
\omega \in \Omega \,\middle|\,
I_{b_\mathrm{SC}}^{(\omega)} < \mathrm{I}_{b_{\mathrm{lim}}}
\right\}
\label{eq:opt_problem_e} \\
& \hspace{5mm}\Omega_2 =
\left\{
\omega \in \Omega \,\middle|\,
\mathrm{I}_{b_{\mathrm{lim}}}
\le I_{b_\mathrm{SC}}^{(\omega)}
< \mathrm{I}_{b_{\mathrm{lim}}} + \nu
\right\}
\label{eq:opt_problem_f} \\
& \hspace{5mm}\Omega_3 =
\left\{
\omega \in \Omega \,\middle|\,
\mathrm{I}_{b_{\mathrm{lim}}} + \nu
\le I_{b_\mathrm{SC}}^{(\omega)}
\right\}
\label{eq:opt_problem_g}
\end{align}
\end{subequations}
where \eqref{eq:opt_problem_b} and \eqref{eq:opt_problem_e} ensure accurate classification of all operating points whose SCC levels are below the limit. A positive parameter $\nu$ is introduced to define regions $\Omega_2$ and $\Omega_3$ (as shown in \eqref{eq:opt_problem_f}-\eqref{eq:opt_problem_g}), such that all samples in $\Omega_3$ are correctly classified, while any misclassified samples are confined to $\Omega_2$. Therefore, $\nu$ should be set to its smallest feasible value to guarantee that the approximation meets the desired accuracy.

The aforementioned classification is merely used to train the coefficients in \eqref{eq:SCC_constraints_linearlized} so as to approximate the actual SCC, and thus acts as an offline preprocessing module for the pricing model. Accordingly, $\mathcal{K}$ are not written in \textit{italics}. Details of the training workflow used in this paper are provided in \cite{Code}.

\vspace{-0.2cm}
\subsection{Existing Schemes for Pricing SCC Services}\label{Existing Schemes for Pricing SCC}
Two previously proposed methods for computing prices of SCC are described next \cite{chu2024pricing}: 

\subsubsection{Dispatchable Pricing}
This method is based on relaxing the binary commitment decisions of SGs to calculate shadow prices. However, this implies that the bilinear term $\eta_{m}$ in \eqref{eq:SCC_constraints_linearlized} becomes a product of two continuous variables and can no longer be exactly linearized using techniques such as McCormick envelopes. Therefore, $\eta_{m}$ has to be excluded in order to apply the dispatchable pricing, leading to a simplified form of the SCC constraint:
\begin{equation}  \label{eq:SCC_cons_no_eta}
\sum_{g}\mathrm{k}_{bg}u_{g}+\sum_{c}\mathrm{k}_{bc}\alpha_{c}\ge\mathrm{I}_{b_{\mathrm{lim}}}
\end{equation}
which linearly expresses the SCC level, bringing a low computational cost. Nevertheless, this model comes at the expense of physical misrepresentation: the energy and SCC markets may be coupled through unrealistic operating conditions, as unit hard constraints cannot be strictly satisfied. The discarded term $k_{bm}$, which is multiplied by $\eta_{m}$, may be either positive or negative; therefore, the obtained SCC prices may even lead to a violation of system security due to insufficient SCC levels.

\subsubsection{Restricted Pricing}\label{Restricted Pricing}
The restricted pricing method proceeds in three sequential steps:

\textbf{Step 1}: Solve the original SCC‑constrained UC problem \eqref{eq:primal_model}–\eqref{eq:MC_linear}, which yields integer‑valued optimal solutions for variables $u_g^*$ and $\eta_m^*$.

\textbf{Step 2}: Relax the integer constraints on $u_g$ and $\eta_m$ such that they become continuous variables. Introduce equality constraints to fix these relaxed continuous variables to the pre‑computed optimal values from \textbf{Step 1}, i.e., $u_{g}=u_{g}^*$ and $\eta_{m}=\eta_{m}^*$. Let $\lambda_{g,\mathrm{commit}}$ denote the Lagrange multiplier associated with constraint $u_{g}=u_{g}^*$.

\textbf{Step 3}: Solve the dual problem \eqref{eq:dual_model} augmented with the new Lagrange term $\lambda_{g,\mathrm{commit}}(u_g^*-u_g)$ defined in \textbf{Step 2}, and extract the resulting shadow‑price solution $\lambda_{g,\mathrm{commit}}$ and $\lambda^\mathrm{E}$.

This method thus retains the model integrality and complete form of the SCC constraint. Although $u_{g}$ and $\eta_{m}$ are fixed to integer values, they are defined as continuous variables in the pricing stage, which enables the computation of shadow prices.

However, this method renders SCC constraints non-binding in the second-stage problem. Instead, the only non-zero associated shadow price is `$\lambda_{g,\mathrm{commit}}$', which contributes to uplift payments to ensure that necessary units remain online for both energy and grid-stability purposes. The price for all ancillary services that can be classified as `all or nothing', that is, those that are provided solely based on the on/off status of a unit, is effectively bundled into the value of $\lambda_{g,\mathrm{commit}}$. This not only reduces price interpretability but also leads to a key limitation for a classical technology that has regained attention in recent years: synchronous condensers. These assets are valuable for providing SCC but would not be properly remunerated for this service under the `restricted pricing' scheme, since they lack clearly defined SCC price signals, as discussed in Section~\ref{Profitability of Generators under Different Pricing Methods}.

To overcome these pitfalls in SCC pricing, a method based on the P-D formulation is proposed. It aims to determine explicit SCC prices that adequately compensate generators while preserving model integrality and avoiding make-whole payments. A comparative summary of the different pricing schemes is shown in Table \ref{table:pricing-schemes}. Note that since marginal-unit pricing cannot be integrated into the general dual-based market framework, this paper does not examine it in detail.

\begin{table}[t]
\centering
\caption{Main Features of Various Schemes for \\ Pricing Short-Circuit Current Services}
\vspace{-0.3cm}
\setlength{\tabcolsep}{2.2pt}
{
\renewcommand{\arraystretch}{1.2}
\begin{tabular}{lccc}
\toprule
~~~~~~Schemes & UC property & Shadow prices \\
\midrule
Dispatchable pricing   & Relaxed & \(\lambda_{b}^{\mathrm{SCC}}\) \\
Restricted pricing      & Integer    & \(\lambda_{g,\mathrm{commit}}\) (bundled) \\
Primal-dual pricing & Integer    & \(\lambda_{b}^{\mathrm{SCC}}\) \\
\bottomrule
\end{tabular}
}
\label{table:pricing-schemes}
\vspace{-0.6cm}
\end{table}

\vspace{-0.2cm}
\subsection{Non-Convex Electricity Market Equilibrium Principles}\label{Relevant Economic Principles and Market Settings}

Given their importance in pricing within a UC setting, this subsection clarifies the uplift payments arising from non-convex electricity markets, as well as how such payments would be handled to ensure the market reaches equilibrium.

\subsubsection{Uplift Payments}\label{Uplift Payment}
Due to the on/off operation characteristics of thermal units, non-convexities arise, which may result in no exact prices that support the dispatch, since the set prices may not support units' profit-maximizing operating points, and fixed costs may not be covered \cite{gribik2007market}. One way to address this issue is to provide `uplift payments' to compensate units for losses or lost opportunities when following the dispatch.

The `make-whole payments' are part of the aggregate uplift, which guarantee that a unit can recover its offer-based costs not covered by energy and ancillary service revenues, and are computed as `$\mathrm{max} \{0, \mathrm{Costs} - \mathrm{Captured~Revenues} \}$'. Another form of uplift is Lost Opportunity Costs (LOCs), which represent the potential profits forgone by units when following the operator's dispatch. They are defined as the gap between the maximum profit a unit could hypothetically achieve by independently optimizing its operating decisions at the given market prices and the actual profit obtained through the dispatch \cite{hua2016convex}, which is given by \eqref{eq:LOCs}.

However, uplift payments may reduce market transparency and make it hard for potential entrants to assess investment profitability \cite{hua2016convex}. Although Convex Hull Pricing (CHP) addresses this issue by determining prices that minimize aggregate LOCs, it does not necessarily eliminate all uplifts, and the allocation of these costs remains an important issue \cite{ruiz2012pricing}.

\subsubsection{Quasi-Partial Equilibrium}\label{Quasi-Partial Equilibrium}
We consider a single-shot market in which units accept market outcomes as given and cannot revise their submitted offers or independently adjust their operating points after market clearing. In this setting, the P-D formulation can determine prices that ensure no unit incurs financial losses, without requiring additional make-whole payments.

Although some units may still experience LOCs under the P-D prices, as shown in Section~\ref{Potential Uplift Payments for Lost Opportunity Costs}, these do not necessarily undermine the quasi-partial equilibrium \cite{ruiz2012pricing}. Unlike in a competitive partial equilibrium, units are assumed to follow the operator's dispatch rather than independently re-optimize their operating decisions for profits. Thus, as long as their costs are recovered, they would have no financial incentive to reject the market outcomes. In contrast, CHP may not guarantee a quasi-partial equilibrium only through cleared service prices, since make-whole payments may remain \cite{hua2016convex}, requiring a dedicated uplift allocation mechanism.

\section{Primal-Dual Formulation for Pricing Short-Circuit Current Services}\label{Primal-Dual Formulation for Pricing SCC Services}
Here, we first introduce the mathematical foundation, and outline the implementation procedure of the proposed pricing framework. By introducing a generic UC problem subject to SCC constraints, we derive the mathematical formulation of the proposed method, which is ultimately expressed as a Mixed-Integer Linear Program (MILP).

\vspace{-0.2cm}
\subsection{Foundations of Pricing via Optimization} \label{Basic Mathematical Background of Pricing Procedure}
Considering the Linear Program (LP) on the left-hand side of \eqref{eq:primal_dual_final}, its dual counterpart can be derived as shown on the right-hand side.
\begin{equation}\label{eq:primal_dual_final}
\setlength{\arraycolsep}{5pt} 
\renewcommand{\arraystretch}{1.0} 
\begin{array}{@{}llc!{\vline}llc@{}}
\min\limits_{x} && c^\mathrm{T} x 
&
\max\limits_{\mu} && b^\mathrm{T} \mu \\
\mathrm{s.t.}      && Ax \geq b,\, x \geq 0 
&
\mathrm{s.t.}      && A^\mathrm{T} \mu \leq c,\, \mu \geq 0
\end{array}
\end{equation}
where $x \in \mathbb{R}^n$, $c \in \mathbb{R}^n$, $A \in \mathbb{R}^{m \times n}$, $b \in \mathbb{R}^m$ and $\mu \in \mathbb{R}^m$, with the shadow prices included in the dual variable vector $\mu$. 

As both problems in \eqref{eq:primal_dual_final} are linear and convex, the complementary slackness in Karush–Kuhn–Tucker (KKT) conditions is necessary and sufficient for optimality and can be given as:
\begin{equation}\label{eq:complementary_slackness}
0 \leq (c - A^\mathrm{T} \mu) \perp x \geq 0
\quad ; \quad
0 \leq \mu \perp (Ax - b) \geq 0
\end{equation}
where $y \perp z $ indicates complementarity, that is, $y \cdot z = 0$.

By relaxing the complementarity conditions, i.e., imposing soft constraints on \eqref{eq:complementary_slackness} and allowing the dot products therein to be positive while bounding them below a certain threshold, the non-negative slack variables $\epsilon_1$ and $\epsilon_2$ are introduced such that $ \epsilon_1 \geq (c - A^\mathrm{T} \mu)^\mathrm{T} x $ and $ \epsilon_2 \geq (Ax - b)^\mathrm{T} \mu $. Under this relaxation, \eqref{eq:complementary_slackness} can be reformulated as the left-hand side of \eqref{eq:relaxation_complementary_slackness}, which is equivalent to the right-hand side formulation, as illustrated by the arrows.
\begin{equation}\label{eq:relaxation_complementary_slackness}
\setlength{\arraycolsep}{0pt}  
\renewcommand{\arraystretch}{1}  
\begin{array}{@{}l@{}c@{\Rightarrow\hspace{3pt}}l@{}c@{}}
\min\limits_{x,\mu,\epsilon_1,\epsilon_2} & \hspace{-5pt} \epsilon_1 + \epsilon_2
&
\min\limits_{x,\mu} & \hspace{4pt} (c-A^\mathrm{T}\mu)^\mathrm{T}x + (Ax-b)^\mathrm{T}\mu \\
\mathrm{s.t.}         & \begin{aligned}[t]
& \hspace{-15pt} Ax\geq b,\,x\geq0 \\
& \hspace{-15pt} A^\mathrm{T}\mu\leq c,\,\mu\geq0
\end{aligned}
&
\mathrm{s.t.}         & \begin{aligned}[t]
& \hspace{-45pt} Ax\geq b,\,x\geq0 \\
& \hspace{-45pt} A^\mathrm{T}\mu\leq c,\,\mu\geq 0
\end{aligned}
\end{array}
\end{equation}

Note that \eqref{eq:relaxation_complementary_slackness} is eventually equivalent to:
\begin{subequations}\label{eq:math_P_D}
\begin{align}
& \min\limits_{x,\mu} \ \epsilon = c^\mathrm{T} x - b^\mathrm{T} \mu
\label{eq:math_P_D_obj}
\\
& \mathrm{s.t.} \quad Ax \geq b,\, x \geq 0; ~ A^\mathrm{T} \mu \leq c,\, \mu \geq 0
\label{eq:math_P_D_con}
\end{align}
\end{subequations}
where \eqref{eq:math_P_D_obj} minimizes the duality gap formed by the objectives in \eqref{eq:primal_dual_final}, subject to the primal and dual constraints in \eqref{eq:math_P_D_con}.

Strong duality holds for \eqref{eq:primal_dual_final} when $\epsilon=0$. Although formulation \eqref{eq:math_P_D} allows the inclusion of extra constraints to capture problem property and desired solution features, such as binary variables and the linking constraint \eqref{eq:nonnegative_profits_cons} coupling primal and dual variables (e.g., power output and energy prices), these additions typically violate strong duality and yield $\epsilon \geq 0$. Nevertheless, the solution of \eqref{eq:math_P_D} stays as close as possible to that of \eqref{eq:complementary_slackness}, as both primal and dual feasibility conditions are satisfied and the relaxed feasible region is always constructed around the optimum of \eqref{eq:primal_dual_final}.

\subsection{Primal Formulation of SCC-Constrained UC }
Without loss of generality, we consider a system with both SGs and IBR, where a UC is formulated to minimize operating costs while satisfying the SCC requirement at each bus. The single-period UC is formulated below, with the corresponding dual variables assigned to the right-hand side of each constraint.
\begin{subequations}\label{eq:primal_model}
 \begingroup
\begin{alignat}{2}
    &  \displaystyle \min_{V_\mathrm{P}}   
     \sum_g \big( \mathrm{c}_{g}^\mathrm{nl} u_{g} +\mathrm{c}_{g}^\mathrm{m} P_{g}+ C_{g}^\mathrm{st}  \big)   \label{eq:primal_obj}  
\end{alignat}
\endgroup     
$\mathrm{where:}$
\begin{alignat}{3}
& V_\mathrm{P}  = \Bigl\{ \; u_{g}, P_{g}, C_{g}^\mathrm{st}, P_{c}, \eta_{m} \; \Bigl\}     \label{eq:primal_variables} 
\end{alignat}
$\mathrm{subject~to:}$
\begin{alignat}{4}  
    & \sum_{g}P_{g}+ \sum_{c}P_{c} = \mathrm{P}^\mathrm{D}:  (\lambda^{\mathrm{E}}) \label{eq:primal_cons_power_balance} \\
    & u_{g} \mathrm{P}_{g}^\mathrm{min}  \leq P_{g} \leq u_{g}  \mathrm{P}_{g}^\mathrm{max}:  ({\mu}^{\mathrm{min}}_{g}, {\mu}^{\mathrm{max}}_{g}),~ \forall g \label{eq:primal_cons_SG_output}\\ 
    & C_{g}^\mathrm{st} \ge 0:  (\rho_{g}^\mathrm{st}),~ \forall g  \label{eq:primal_cons_st_cost_positive}\\
    & C_{g}^\mathrm{st} \ge (u_{g}-u_{g,0})\mathrm{c}^\mathrm{st}_{g}:  (\sigma_{g}^\mathrm{st}),~ \forall g \label{eq:primal_cons_st_cost_lb}\\
    & 0 \leq P_{c} \leq \alpha_{c}\mathrm{P}^{\mathrm{max}}_{c}:  (\zeta_{c}^\mathrm{min},\zeta_{c}^\mathrm{max}),~ \forall c  \label{eq:primal_cons_wt}\\
    & u_{g} \in \{0,1\} \label{eq:binary_SGs},~ \forall g        \\
    & \sum_{g}\mathrm{k}_{bg}u_{g}\hspace{-0.05cm}+\hspace{-0.1cm}\sum_{c}\mathrm{k}_{bc}\alpha_{c}\hspace{-0.05cm}+\hspace{-0.1cm}\sum_{m}\mathrm{k}_{bm}\eta_{m} \hspace{-0.05cm} \ge \hspace{-0.05cm} \mathrm{I}_{b_{\mathrm{lim}}}: ( \lambda_{b}^{\mathrm{SCC}} ), \forall b~ \label{eq:primal_cons_SCC} \\  
    & \eta_m= \{u_{g_\mathrm{1}} \cdot u_{g_\mathrm{2}} \mid {g_\mathrm{1},g_\mathrm{2}}=m \}    \label{eq:interactions_pair_SGs_1_P_D}         \\ 
    & m\in\mathcal{M}=\{g_{\mathrm{1}},g_{\mathrm{2}}\mid \forall g_{\mathrm{1}}, g_{\mathrm{2}}\in\mathcal{G}\}    \label{eq:interactions_pair_SGs_2_P_D}   \\
    &  \mathrm{McCormick~envelopes~for~linearizing~}\eta_m, ~\forall m \label{eq:primal_Mccormick_env}        
\end{alignat}   
\end{subequations}
where the system operating costs are defined as \eqref{eq:primal_obj}, including the no-load, marginal generation, and startup costs of SGs, while the energy supplied by IBR is assumed to be cost-free. Eq.~\eqref{eq:primal_variables} states primal variables that follow constraints: Supply-demand power balance \eqref{eq:primal_cons_power_balance}; Generation limits for SGs \eqref{eq:primal_cons_SG_output}; Startup costs \eqref{eq:primal_cons_st_cost_positive}-\eqref{eq:primal_cons_st_cost_lb} that account for the initial status `$u_{g,0}$'; Generation limits for IBR \eqref{eq:primal_cons_wt}; Enforcement of the binary property of UC \eqref{eq:binary_SGs}; SCC constraint for bus $b$ \eqref{eq:primal_cons_SCC}; Auxiliary equations for linearizing $\eta_{m}$ (the product of two binary variables) via \eqref{eq:primal_Mccormick_env} are given as \cite{urban2021mccormick}:
\begin{subequations} \label{eq:MC_linear}
\begin{align}
& \eta_{m}\leq u_{g_\mathrm{1}}: (\gamma_{m,1}^{\mathrm{max}}),~\forall m \label{eq:MC_linear_1} \\
& \eta_{m}\leq u_{g_\mathrm{2}}: (\gamma_{m,2}^{\mathrm{max}}),~\forall m \label{eq:MC_linear_2} \\
& \eta_{m} \geq u_{g_\mathrm{1}}+u_{g_\mathrm{2}}-1: (\gamma_{m,1}^{\mathrm{min}}),~\forall m \label{eq:MC_linear_3} \\
& \eta_{m} \in \{0, 1\},~\forall m \label{eq:MC_linear_4}
\end{align}
\end{subequations}
Consequently, the primal is reformulated to an MILP, consisting of \eqref {eq:primal_model} and \eqref {eq:MC_linear}.

\vspace{-0.3cm}
\subsection{Dual Formulation of Relaxed SCC-Constrained UC}
Due to integer variables in the primal problem, its dual formulation requires relaxing integrality constraints:
\begin{subequations}  \label{eq:integer_relax}
\begin{align}
& 0 \leq u_g \leq 1: (\psi_{g}^{\mathrm{min}}, \psi_{g}^{\mathrm{max}}),~ \forall g \\
& \eta_{m} \geq 0: (\gamma_{m,2}^{\mathrm{min}}),~ \forall m  \label{eq:MC_linear_5}
\end{align}
\end{subequations}
where the upper bound of $\eta_{m}$ is still determined by $u_{g_\mathrm{1}}$ or $u_{g_\mathrm{2}}$ and is bounded by 1, as in \eqref{eq:MC_linear_1} and \eqref{eq:MC_linear_2}.

By relaxing the binary variables, the MILP is converted into an LP, from which the dual problem in \eqref{eq:dual_model} can be derived:
\begin{subequations}\label{eq:dual_model}
\begin{alignat}{2}
    &  \displaystyle \max_{V_\mathrm{D}}  ~ \mathrm{P}^\mathrm{D}\lambda^{\mathrm{E}}+ \hspace{-0.1cm} \sum_b(\mathrm{I}_{b_{\mathrm{lim}}} \hspace{-0.1cm} - \hspace{-0.1cm} \sum_{c}\mathrm{k}_{bc}\alpha_{c})\lambda_{b}^{\mathrm{SCC}} \hspace{-0.1cm} - \hspace{-0.1cm} \sum_c\alpha_{c}\mathrm{P}^{\mathrm{max}}_{c}\zeta_{c}^\mathrm{max} \nonumber \\
     &  -  \sum_{g}\psi_{g}^{\mathrm{max}} -\sum_{m}\gamma_{m,1}^{\mathrm{min}} - \sum_{g}  u_{g,0}\mathrm{c}_{g}^{\mathrm{st}}\sigma_{g}^{\mathrm{st}} 
    \label{eq:dual_obj}  
\end{alignat}
$\mathrm{where:}$
\begin{alignat}{2}
& V_\mathrm{D}  = \Bigl\{ \; \lambda^{\mathrm{E}},\lambda_{b}^{\mathrm{SCC}},\zeta_{c}^\mathrm{max},\psi_{g}^{\mathrm{max}}, \gamma_{m,1}^{\mathrm{max}}, \gamma_{m,2}^{\mathrm{max}}, \nonumber \\  
& \qquad \qquad  \gamma_{m,1}^{\mathrm{min}}, \mu^{\mathrm{max}}_{g}, \mu^{\mathrm{min}}_{g}, \sigma_{g}^\mathrm{st}  \; \Bigl\}    \label{eq:dual_variables} 
\end{alignat}
$\mathrm{subject~to:}$
\allowdisplaybreaks
\begin{alignat}{2}
    & \mathrm{c}_{g}^\mathrm{nl} -\sum_b\mathrm{k}_{bg}\lambda_{b}^{\mathrm{SCC}}-\mathrm{P}_{g}^{\mathrm{max}}\mu_{g}^{\mathrm{max}}+\mathrm{P}_{g}^{\mathrm{min}}\mu_{g}^{\mathrm{min}} +\mathrm{c}_{g}^{\mathrm{st}}\sigma_{g}^{\mathrm{st}} \nonumber \\
    &  +h_g(\gamma_{m,1}^{\mathrm{max}}, \gamma_{m,2}^{\mathrm{max}}, \gamma_{m,1}^{\mathrm{min}})  +\psi_{g}^{\mathrm{max}} \ge 0,~\forall g  \label{eq:dual_cons_binary_1} \\
    & \mathrm{c}_{g}^\mathrm{m}-\lambda^{\mathrm{E}}+\mu^{\mathrm{max}}_{g}-\mu^{\mathrm{min}}_{g} \ge 0,~\forall g \label{eq:dual_cons_Pg_2} \\
   & 1-\sigma_{g}^{\mathrm{st}} \ge 0,~ \forall g \label{eq:dual_cons_Cst} \\
    & -\lambda^{\mathrm{E}}+\zeta^{\mathrm{max}}_{c} \ge 0,~\forall c \label{eq:dual_cons_Pc}\\
    & \gamma_{m,1}^{\mathrm{max}} + \gamma_{m,2}^{\mathrm{max}} - \hspace{-0.05cm} \gamma_{m,1}^{\mathrm{min}} - \sum_b\mathrm{k}_{bm}\lambda_{b}^{\mathrm{SCC}}  \geq 0,~\forall m  \label{eq:dual_eta} \\ 
    & \{V_\mathrm{D} | V_\mathrm{D} \neq \lambda^{\mathrm{E}} \}   \in \mathbb{R}_+,~\forall b, g, c, m \label{eq:dual_var_nonnegative}
\end{alignat}
\end{subequations}
where \eqref{eq:dual_obj} is the objective function of the dual, and \eqref{eq:dual_variables} shows the associated dual variables. The correspondence between dual constraints and primal variables is: \eqref{eq:dual_cons_binary_1}$\leftrightarrow$$u_{g}$, \eqref{eq:dual_cons_Pg_2}$\leftrightarrow$$P_{g}$, \eqref{eq:dual_cons_Cst}$\leftrightarrow$$C_{g}^\mathrm{st}$, \eqref{eq:dual_cons_Pc}$\leftrightarrow$$P_{c}$, \eqref{eq:dual_eta}$\leftrightarrow$$\eta_{m}$. Constraint \eqref{eq:dual_var_nonnegative} enforces the non-negativity of dual variables related to inequality constraints, i.e., all dual variables except \(\lambda^\mathrm{E}\) are nonnegative. $h_g(\gamma_{m,1}^{\mathrm{max}}, \gamma_{m,2}^{\mathrm{max}}, \gamma_{m,1}^{\mathrm{min}})$ is the dual term associated with $\eta_{m}$ in McCormick envelopes \eqref{eq:primal_Mccormick_env}, which is detailed in Appendix~\ref{Dual constraints for UC states in McCormick envelopes}. Given that IBR have limited overload capacity \cite{chu2021short}, their SCC contributions are substantially smaller than SGs. However, such volumes are non-negligible and in fact enter the term $(\mathrm{I}_{b_{\mathrm{lim}}} -  \sum_{c}\mathrm{k}_{bc}\alpha_{c})\lambda_{b}^{\mathrm{SCC}}$ in \eqref{eq:dual_obj} to represent the net SCC requirement at bus $b$, which is priced at $\lambda_{b}^{\mathrm{SCC}}$ and fulfilled by thermal units.

\subsection{Primal-Dual Formulation for Pricing SCC Services}\label{Primal-Dual Formulation}
This subsection gives the compact form of the final P-D pricing and the necessary linearization procedure.

\subsubsection{Complete Formulation}

The objective of the proposed P-D formulation is to determine $\lambda^\mathrm{E}$ and $\lambda_b^\mathrm{SCC}$ under non-convexities, rather than commitment prices and uplift payments. That is, prices that provide proper incentives to generators and ensure that each dispatched thermal unit recovers its costs through market participation. Accordingly, as discussed in Section~\ref{Basic Mathematical Background of Pricing Procedure}, the following non-negative profit constraint can be imposed on each thermal unit:
\begin{equation}\label{eq:nonnegative_profits_cons}
\begin{alignedat}{2}
    & \underbrace{\lambda^\mathrm{E} P_g}_{\mathrm{Energy~revenue}} 
   \hspace{-0.2cm} + \underbrace{\sum_b \lambda_b^\mathrm{SCC} \mathrm{k}_{bg} u_g
    + \sum_b \sum_{ \{m | g \in m \} } \lambda_b^\mathrm{SCC} \mathrm{k}_{bm} \eta_m}_{\mathrm{SCC~revenue}}  \\
    & - ( \underbrace{\mathrm{c}_{g}^\mathrm{nl} u_{g} + \mathrm{c}_{g}^\mathrm{m} P_{g}
    + C_{g}^\mathrm{st} )}_{\mathrm{Operating~cost}}
    \geq 0, \quad \forall g
\end{alignedat}
\end{equation}

Note that the non-negative profit requirement does not apply to IBR, since these assets are modeled as cost-free in this work. Should this be required, their profit constraint would follow the structure of \eqref{eq:nonnegative_profits_cons} but without fixed-cost terms, as analyzed in Section~\ref{Overview of Market Settlement}.

Finally, the P-D formulation for optimizing market clearing prices is compactly expressed as follows:
\begin{subequations}\label{eq:final_pricing}
 \begingroup
\begin{alignat}{2}
    &  \displaystyle \min_{V}~\eqref{eq:primal_obj}-\eqref{eq:dual_obj}   \label{eq:final_pricing_obj} 
\end{alignat}
\endgroup
$\mathrm{where:}$
\begin{alignat}{3}
&   V= \Bigl\{ \; V_\mathrm{P}~ \eqref{eq:primal_variables}, V_\mathrm{D}~ \eqref{eq:dual_variables} \;\Bigl\}     \label{eq:final_pricing_variables} 
\end{alignat}
$\mathrm{subject~to:}$
\begin{alignat}{4}  
    & \mathrm{Primal~constraints:}~ \eqref{eq:primal_cons_power_balance}\mathrm{-}\eqref{eq:MC_linear} \\ 
    & \mathrm{Dual~constraints:}~ \eqref{eq:dual_cons_binary_1}\mathrm{-}\eqref{eq:dual_var_nonnegative} \\
    & \mathrm{Non\text{-}negative~profit~constraint:}~\eqref{eq:nonnegative_profits_cons}
\end{alignat}  
\end{subequations}
where \eqref{eq:final_pricing_obj} denotes the gap between the primal and dual objective values after incorporating the linking constraint \eqref{eq:nonnegative_profits_cons}, referred to as the P-D gap.

It is worth emphasizing that the UC schedule is determined simultaneously with prices in the P-D formulation. By contrast, under the restricted method, \(\lambda_{g,\mathrm{commit}}\) stems from the artificial equality constraints \(u_g = u_g^*\). Introducing such constraints and multipliers into the P-D formulation would re‑bundle SCC value with other unit‑status‑dependent services and eliminate the explicit locational SCC shadow prices $\lambda_b^\mathrm{SCC}$ we aim to obtain. Instead, the proposed framework computes SCC value directly from the shadow prices of nodal SCC constraints and enforces revenue adequacy via constraint \eqref{eq:nonnegative_profits_cons}.

\subsubsection{Linearization for Terms in Profit Constraints}

Note that \eqref{eq:nonnegative_profits_cons} incorporates the nonlinear terms `$\lambda^\mathrm{E} P_g$', `$\lambda_b^\mathrm{SCC} \mathrm{k}_{bg} u_g$' and `$\lambda_b^\mathrm{SCC} \mathrm{k}_{bm} \eta_m$', which typically make the model computationally challenging. To improve computational tractability, these terms should first be properly linearized. Specifically, the bilinear term `$\lambda^\mathrm{E} P_g$' is linearized using the binary expansion approach described below \cite{pereira2005strategic}:
\begin{subequations}\label{eq:nonnegative_profit}
\begin{align}
& P_g = u_g (\mathrm{P}_g^\mathrm{min} + \sum_{n=1}^\mathrm{N} 2^{n-1} \Delta \mathrm{P}_g s_n \label{eq:binary_expansion_1} ) \\
& \Delta \mathrm{P}_g = \frac{\mathrm{P}_g^\mathrm{max} - \mathrm{P}_g^\mathrm{min}}{2^\mathrm{N}-1} \label{eq:binary_expansion_2}\\
& \lambda^\mathrm{E} P_g = \lambda^\mathrm{E} u_g (\mathrm{P}_g^\mathrm{min} + \sum_{n=1}^\mathrm{N} 2^{n-1} \Delta \mathrm{P}_g s_n )\label{eq:binary_expansion_3}\\
& v_n = \lambda^\mathrm{E} s_n  \label{eq:binary_expansion_4}\\
& 0 \leq \lambda^\mathrm{E} - v_n \leq \lambda^\mathrm{E,max} (1-s_n),~0 \leq v_n \leq \lambda^\mathrm{E,max} s_n \label{eq:binary_expansion_5}
\end{align}
\end{subequations}
where \eqref{eq:binary_expansion_1} and \eqref{eq:binary_expansion_2} express $P_g$ as a sum of auxiliary binary variables $s_n$, thereby rewriting `$\lambda^\mathrm{E} P_g$' as \eqref{eq:binary_expansion_3}. The product in \eqref{eq:binary_expansion_4} is then linearized through \eqref{eq:binary_expansion_5}, with the energy price restricted to be non-negative, so as to keep market agents financially whole. Note that the auxiliary variable $v_n$ in \eqref{eq:binary_expansion_4}-\eqref{eq:binary_expansion_5} serves only for linearizing the bilinear product and is not an essential part of the physical power‑output decomposition of $P_g$ in \eqref{eq:binary_expansion_1}. The parameter $\lambda^\mathrm{E,max}$ is a sufficiently large constant chosen so that \eqref{eq:binary_expansion_5} remains non-binding at optimality. Since $\lambda^\mathrm{E}$ and $v_n$ denote energy prices, $\lambda^\mathrm{E,max}$ can be set as $ 2 \cdot \mathrm{max}\{\mathrm{c}^\mathrm{m}_g~|~\forall g \}$. The parameter $\mathrm{N}$ is the number of segments in the discretized power output range, which allows \eqref{eq:binary_expansion_1} to more closely approximate the continuous variable $P_g$ with a larger value of $\mathrm{N}$. To this end, the product term `$\lambda^\mathrm{E} u_g s_n$' in \eqref{eq:binary_expansion_3} can be further expressed as `$u_gv_n$' based on \eqref{eq:binary_expansion_4}; this product of a discrete variable and a continuous variable can be linearized in a similar manner using \eqref{eq:binary_expansion_5}, so it is not elaborated here.

The products of continuous and binary variables, i.e., `$\lambda_b^\mathrm{SCC} \mathrm{k}_{bg} u_g$' and `$\lambda_b^\mathrm{SCC} \mathrm{k}_{bm} \eta_m$', can be similarly linearized by imposing suitable bounds on $\lambda_b^\mathrm{SCC}$. Specifically, $\lambda_b^\mathrm{SCC}$ can be taken as twice the sum of no-load and startup costs, thereby ensuring adequate remuneration for thermal units committed for SCC provision. Hence, the following constraint is included in the model: $ 0 \leq \lambda_b^\mathrm{SCC} \leq 2 \cdot \mathrm{max} \{ \mathrm{c}_g^\mathrm{nl} + \mathrm{c}_g^\mathrm{st}~|~\forall g  \} $.

\begin{figure}
    \centering
    \includegraphics[width=1\linewidth]{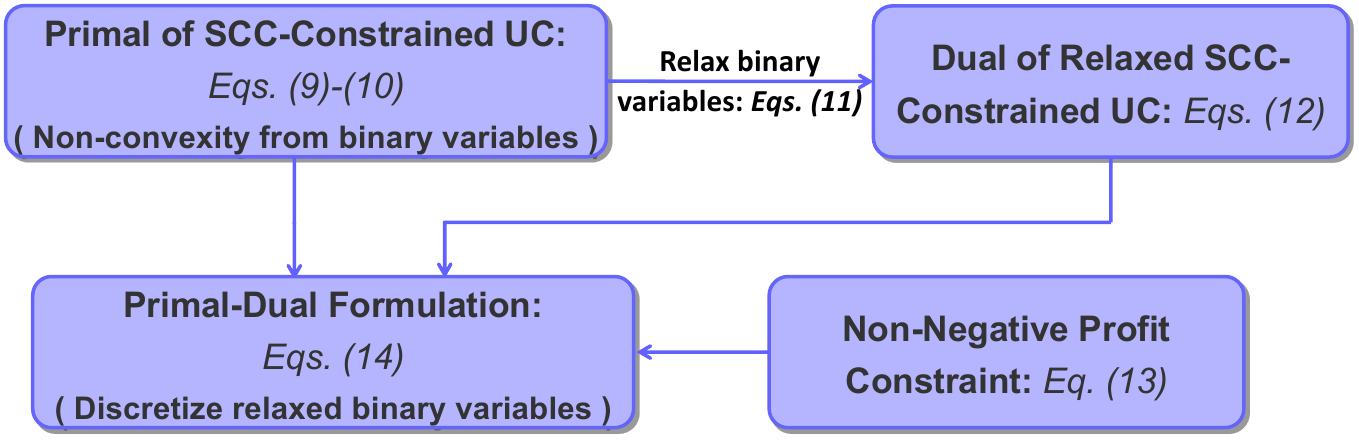}
                \vspace{-0.6cm}
    \caption{Flow chart of P-D formulation for pricing SCC services.}     \label{fig:primal_dual_flowchart}
                    \vspace{-0.2cm}
\end{figure}

\begin{figure}
    \centering
    \includegraphics[width=0.9\linewidth]{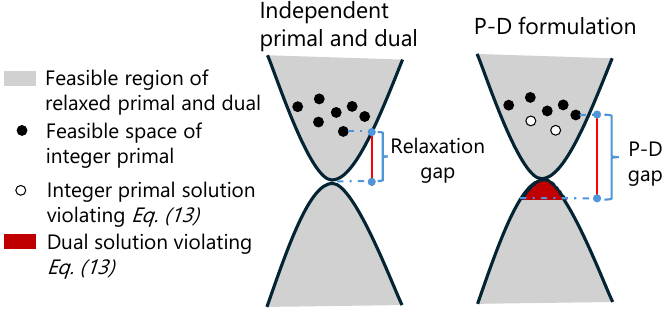}
            \vspace{-0.3cm}
    \caption{Primal and dual solution comparison with and without constraint \eqref{eq:nonnegative_profits_cons}. Note that the relaxation gap may nearly coincide with the P-D gap, owing to the degrees of freedom across primal and dual variables, as the case in Section~\ref{Horizon-Aggregated Revenue Adequacy}.}
    \label{fig:primal_dual_diagram}
    \vspace{-0.3cm}
\end{figure}

After linearization, the resulting optimization becomes an MILP, as described in Fig.~\ref{fig:primal_dual_flowchart}. It is noteworthy that solving problem \eqref{eq:final_pricing} without the non-negative profit constraint \eqref{eq:nonnegative_profits_cons} is essentially equivalent to separately solving the primal and dual problems (shown in the left-hand side of Fig.~\ref{fig:primal_dual_diagram}), since no explicit objective function or constraint links the variables of the two problems. In this case, the relaxation gap to be minimized is zero when the primal is relaxed and becomes positive when integrality constraints are enforced. When constraint \eqref{eq:nonnegative_profits_cons} is included, the operating points of units in the primal and the service prices from the dual can be jointly optimized. Consequently, the objective of formulation \eqref{eq:final_pricing} (shown in the right-hand side of Fig.~\ref{fig:primal_dual_diagram}) becomes the minimization of the P-D gap, allowing service prices to accurately capture both energy and SCC contributions of units to the system and ensuring non-negative profits for all units.

\section{Case Studies}\label{Case Studies}
Before presenting the formal case study analysis, it is necessary to clarify the adopted system configuration and the case study design. In this work, each dispatch interval is set to one hour, so energy volumes (MWh) equal power (MW) multiplied by one hour.
   
  \vspace{-0.4cm}
\subsection{Experimental Design}

\subsubsection{Test System Setting} \label{Test System Setting}

\begin{figure}[t]
\centering
\includegraphics[width=0.7\columnwidth]{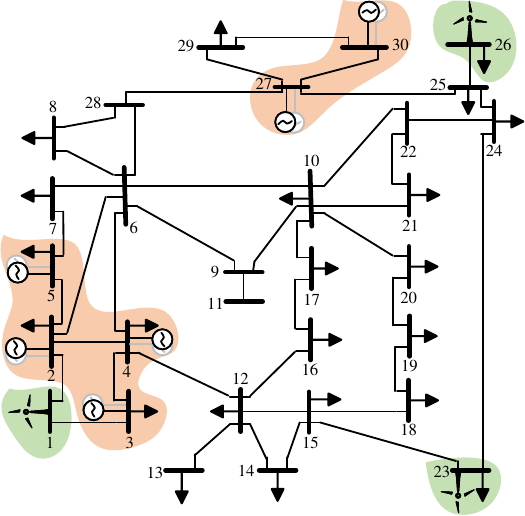}
     \vspace{-0.2cm}
\caption{Modified IEEE 30-bus system.}
\label{ieee_system}
     \vspace{-0.5cm}
\end{figure}  

Case studies are conducted on a modified IEEE 30-bus system (as depicted in Fig.~\ref{ieee_system}) to test the SCC pricing schemes. The IBR, wind turbines, are placed at buses \{1, 23, 26\}, while SGs are located at buses \{2, 3, 4, 5, 27, 30\}, with each bus hosting two SGs. The notation for units is defined as follows: $g_1\text{-}b2$ and $g_2\text{-}b2$ denote the first and second SGs located at bus 2, respectively, while $2g\text{-}b2$ represents both of them. The total capacity of wind power is $750~\mathrm{MW}$. The parameters of SGs are listed in Table \ref{table:SGs_para}. The base power is $S_\mathrm{B}=100~\mathrm{MVA}$. The short-circuit injection of the IBR, $\mathrm{I}_c$, equals $1.00~\mathrm{p.u}$. A nominal voltage of $0.95~\mathrm{p.u}$. is adopted for conservativeness, and $\nu=0.1$ provides the intended approximation accuracy for the SCC constraints. The SCC threshold $\mathrm{I}_{b_{\mathrm{lim}}}$ is set to $5~\mathrm{p.u}$. The approximate SCC coefficients $\mathcal{K}$ were computed in \cite{wang2026imperfect} and used in this paper. The parameter used to linearize bilinear terms in \eqref{eq:nonnegative_profit} is set as `$\mathrm{N}=20$' for an accurate solution, as analyzed in Section~\ref{On Number of Segments}.

As discussed in Section~\ref{Primal-Dual Formulation}, the problem to be solved is a standard MILP, which can be efficiently handled by off-the-shelf optimization solvers. Accordingly, all simulations in this work were run using \texttt{Julia-JuMP} and solver \texttt{Gurobi-12.0.1} on a MacBook Air (M1, 2020). No custom‑tailored solution algorithms or heuristics have been developed in this paper. The code used for case studies is publicly available in the repository \cite{Code}.


\subsubsection{Case Study Arrangement} 
Given that energy demand levels are closely correlated with commitment status of SGs and thereby exert an indirect effect on SCC supply, it is relevant to investigate market outcomes under diverse load conditions. Therefore, the case studies are structured as follows: In a single-period setting, Section~\ref{Single-Period SCC Service Pricing} first clarifies the effect of non-negative profit constraints on the problem solution, and then demonstrates that security constraints are required for SCC-risky buses. It subsequently investigates the service prices and generator profits obtained through the proposed P–D method, and provides comparative analyses against the dispatchable and restricted methods. While Section~\ref{Multi-Period SCC Service Pricing} extends the established pricing framework to a 24-hour time horizon to examine the full-day market clearing process. Section~\ref{Overview of Market Settlement} clarifies the market settlement. Section~\ref{Sensitivity of Pricing to Binary Expansion Errors} analyzes the impact of parameter settings during the linearization process on pricing outcomes. Section~\ref{Applicability of the Primal-Dual Pricing for SCC Services} discusses the applicability of the proposed market framework.


\begin{table}[t]
\centering
\caption{Operating Parameters of Synchronous Generators}
\vspace{-0.3cm}
\setlength{\tabcolsep}{5pt}
{\fontsize{8pt}{12pt}\selectfont
\begin{tabular}{lcccccc}
\toprule
Bus & 2 & 3 & 4 & 5 & 27 & 30 \\
\midrule
\(\mathrm{c}_{g}^\mathrm{nl}\) (\texteuro/h) & 1,743 & 1,501 & 1,376 & 1,093 & 990 & 857 \\
\(\mathrm{c}_{g_1}^\mathrm{m}\) (\texteuro/MWh) & 6.20 & 7.10 & 10.47 & 12.28 & 13.53 & 15.36 \\
\(\mathrm{c}_{g_2}^\mathrm{m}\) (\texteuro/MWh) & 7.07 & 8.72 & 11.49 & 12.84 & 14.60 & 15.02 \\
\(\mathrm{c}_g^\mathrm{st}\) (\texteuro/h) & 20,000 & 12,500 & 9,250 & 7,200 & 5,500 & 3,100 \\
\(\mathrm{P}_g^\mathrm{min}\) (MW) & 658 & 576 & 302 & 133 & 130 & 58 \\
\(\mathrm{P}_g^\mathrm{max}\) (MW) & 1,317 & 1,152 & 756 & 667 & 650 & 576 \\
\(u_{g,0}\) & 1 & 1 & 1 & 0 & 0 & 0 \\
\bottomrule
\end{tabular}
}
\label{table:SGs_para}
\vspace{-0.3cm}
\end{table}

\begin{table}[t]
\centering
\caption{Solution of Single-Period Primal-Dual Formulation under Various Load Levels}
\vspace{-0.3cm}
\setlength{\tabcolsep}{4pt}
{\fontsize{7.5pt}{14pt}\selectfont
\begin{tabular}{l!{\vrule width 0.5pt}l@{\hspace{2pt}}cccccc}
\toprule
\multicolumn{2}{c}{Energy demand (GWh)} & 4.0 & 4.8 & 5.6 & 6.4 & 7.2 & 8.0 \\
\midrule
\multirow{2}{*}{w/o \eqref{eq:nonnegative_profits_cons}} 
& Primal obj. (k\texteuro) & $\mathrm{45.42}$ & $\mathrm{51.08}$ & $\mathrm{57.34}$ & $\mathrm{64.09}$ & $\mathrm{72.60}$ & $\mathrm{83.73}$ \\
& Dual obj. (k\texteuro) & \cellcolor{gray!20}{$\mathrm{43.64}$} & \cellcolor{gray!20}{$\mathrm{49.36}$} & \cellcolor{gray!20}{$\mathrm{55.76}$} & \cellcolor{gray!20}{$\mathrm{63.22}$} & \cellcolor{gray!20}{$\mathrm{72.14}$} & \cellcolor{gray!20}{$\mathrm{83.17}$} \\
\midrule
\multirow{2}{*}{with \eqref{eq:nonnegative_profits_cons}} 
& Primal obj. (k\texteuro) & $\mathrm{45.42}$ & $\mathrm{51.08}$ & $\mathrm{57.34}$ & $\mathrm{64.09}$ & $\mathrm{72.60}$ & $\mathrm{83.73}$ \\
& Dual obj. (k\texteuro) & $\mathrm{43.27}$ & $\mathrm{49.09}$ & $\mathrm{55.28}$ & $\mathrm{63.02}$ & $\mathrm{72.14}$ & $\mathrm{83.17}$ \\
\bottomrule
\end{tabular}
}
\label{table:solving_time_gap}
\vspace{-0.4cm}
\end{table}

\subsection{Single-Period SCC Service Pricing and Market Outcomes}\label{Single-Period SCC Service Pricing}
The solution of the P–D formulation is listed in Table~\ref{table:solving_time_gap}, which reports the impact of the non-negative profit constraint \eqref{eq:nonnegative_profits_cons} on the objective values. The gray region represents the dual solution without enforcing constraint \eqref{eq:nonnegative_profits_cons}, which is equivalent to that of the relaxed primal. Due to the non-convexity introduced by commitment variables, the primal solution deviates from the relaxed one. After introducing constraint \eqref{eq:nonnegative_profits_cons}, the primal solution remains unchanged in this case, while the dual objective value would slightly decrease. These results indicate that the proposed P–D method is able to ensure that generators fully recover their costs through joint optimization of energy and SCC prices.

It is worth noting that a non-zero P-D gap persists regardless of whether the non-negative profit constraint is enforced. This gap fundamentally arises from the non-convex nature of the UC problem, which includes binary commitment variables and SCC constraints, and generally exists. Importantly, constraint \eqref{eq:nonnegative_profits_cons} is not designed to eliminate the P-D gap; rather, at the cost of a slightly increased gap (as in Tables~\ref{table:solving_time_gap} and \ref{table:solving_time_gap_multiperiod}), it ensures that the calculated prices satisfy the revenue-adequacy requirement.

\begin{figure}[!t]
\centering
\begin{tikzpicture}
\begin{axis}[
    width=1.05\linewidth, height=4cm,
    xlabel={\scriptsize Bus},
    ylabel={\scriptsize SCC (p.u.)}, 
    ylabel style={align=center, yshift=-0.1cm},
    label style={font=\scriptsize}, 
    tick label style={font=\scriptsize},
    xmin=1, xmax=11,
    ymin=2, ymax=15,
    xtick={1,2,3,4,5,6,7,8,9,10,11},
    xticklabels={$11$,$13$,$14$,$18$,$19$,$20$,$25$,$26$,$27$,$29$,$30$},
    ytick={2,5,10,15},
    grid=both, grid style={dotted, gray!60},
    axis lines=box,
    legend image post style={xscale=0.4},
]

\addplot[red!40, thin] coordinates {
(1,4.4581) (2,4.6610) (3,4.9748) (4,4.9403) (5,4.9374) (6,4.9990) (7,4.9244) (8,3.8751) (9,4.9190) (10,3.4258) (11,2.5638)
};
\addplot[red, thin] coordinates {
(1,6.8080) (2,7.5407) (3,8.0535) (4,7.9861) (5,7.9928) (6,7.8843) (7,8.8205) (8,4.4056) (9,10.6920) (10,7.5953) (11,8.8111)
};
\addplot[fill=red!20, no markers, draw=none, opacity=0.5] coordinates {
(1,4.4581) (2,4.6610) (3,4.9748) (4,4.9403) (5,4.9374) (6,4.9990) (7,4.9244) (8,3.8751) (9,4.9190) (10,3.4258) (11,2.5638)
(11,8.8111) (10,7.5953) (9,10.6920) (8,4.4056) (7,8.8205) (6,7.8843) (5,7.9928) (4,7.9861) (3,8.0535) (2,7.5407) (1,6.8080)
} -- cycle;

\addplot[blue!40, thin] coordinates {
(1,6.2485)(2,7.0051)(3,7.4757)(4,7.4512)(5,7.4510)(6,7.4780)(7,9.1946)(8,5.0013)(9,12.1332)(10,7.4352)(11,7.5633)
};
\addplot[blue, thin] coordinates {
(1,7.0478)(2,7.9370)(3,8.4443)(4,8.3483)(5,8.3852)(6,8.3817)(7,9.7091)(8,5.0246)(9,12.6488)(10,8.4501)(11,8.9219)
};
\addplot[fill=blue!20, no markers, draw=none, opacity=0.5] coordinates {
(1,6.2485) (2,7.0051) (3,7.4757) (4,7.4512) (5,7.4510) (6,7.4780) (7,9.1946) (8,5.0013) (9,12.1332) (10,7.4352) (11,7.5633)
(11,8.9219) (10,8.4501) (9,12.6488) (8,5.0246) (7,9.7091) (6,8.3817) (5,8.3852) (4,8.3483) (3,8.4443) (2,7.9370) (1,7.0478)
} -- cycle;

\addplot[green!70, dotted, thick, mark=none] coordinates {
(1,4.4761) (2,4.7591) (3,5.0305) (4,4.9804) (5,4.9819) (6,5.0509) (7,4.9677) (8,3.8864) (9,4.9541) (10,3.4383) (11,2.5377)
};
\addplot[orange!70, dotted, thick, mark=none] coordinates {
(1,4.9768) (2,5.3845) (3,5.7085) (4,5.6283) (5,5.6328) (6,5.7392) (7,5.5545) (8,3.9396) (9,5.5617) (10,3.5979) (11,2.6358)
};
\addplot[purple!60, dotted, thick, mark=none] coordinates {
(1,4.9366) (2,5.2485) (3,5.6183) (4,5.5507) (5,5.5576) (6,5.6560) (7,5.4781) (8,3.9230) (9,5.4976) (10,3.5690) (11,2.6483)
};
\addplot[cyan!70, dotted, thick, mark=none] coordinates {
(1,5.8831) (2,6.4441) (3,6.8705) (4,6.7590) (5,6.7615) (6,6.8589) (7,6.4483) (8,4.0148) (9,6.6300) (10,3.7705) (11,2.6965)
};
\addplot[teal!70, dotted, thick, mark=none] coordinates {
(1,5.8831) (2,6.4441) (3,6.8705) (4,6.7590) (5,6.7615) (6,6.8589) (7,6.4483) (8,4.0148) (9,6.6300) (10,3.7705) (11,2.6965)
};
\addplot[brown!70, dotted, thick, mark=none] coordinates {
(1,6.8080) (2,7.5407) (3,8.0535) (4,7.9861) (5,7.9928) (6,7.8843) (7,8.8205) (8,4.4056) (9,10.6920) (10,7.5953) (11,8.8111)
};

\fill[red!20] (1.5,14) rectangle (2,13.1);
\node[font=\scriptsize, right] at (2,13.5) {w/o SCC constraints};

\fill[blue!20] (1.5,12.5) rectangle (2,11.6);
\node[font=\scriptsize, right] at (2,12) {with SCC constraints};
\draw[red!40, thick]    (1.5,10.5) -- (1.7,10.5); \draw[blue!40, thick]    (1.75,10.5) -- (1.95,10.5); \node[font=\scriptsize, right] at (2.05,10.5) {Energy demand $=$ 4.0 GWh};
\draw[red, thick]      (1.5,9.2) -- (1.7,9.2); \draw[blue, thick]      (1.75,9.2) -- (1.95,9.2);  \node[font=\scriptsize, right] at (2.05,9.2) {Energy demand $=$ 8.0 GWh};

\end{axis}
\end{tikzpicture}
\vspace{-0.8cm}
\caption{Range of SCC at risky buses with energy demand from 4.0 to 8.0 GWh. The dotted data points represent the unconstrained SCC levels under various load conditions. SCC threshold $\mathrm{I}_{b_{\mathrm{lim}}} = 5~\mathrm{p.u.}$ The SCC at remaining buses is not shown as they meet the security requirement in each case.}
\label{fig:scc_comparison}
\vspace{-0.4cm}
\end{figure}

\subsubsection{SCC Improvement at Risky Buses}\label{SCC Improvement at Risk Buses}
In UC cases without SCC constraints, insufficient SCC occurs at multiple buses, as indicated by the red region in Fig.~\ref{fig:scc_comparison}. At a demand level of 4.0 GWh (the lower boundary of the red region), the SCC levels at buses \{11, 13, 14, 18, 19, 20, 25, 26, 27, 29, 30\} fall below the security threshold. As energy demand increases, more SGs are committed and passively contribute SCC, enabling most buses to satisfy the required level. However, even when multiple SGs are online to serve a demand of 8.0 GWh (the upper boundary of the red region), bus 26 still fails to meet the security requirement due to its inherent SCC limitation imposed by system impedance. This finding highlights the necessity of explicit constraints that allow SGs to actively regulate system impedance to ensure adequate SCC.

With SCC constraints enforced, all buses maintain SCC above the minimum limit (the blue region in Fig.~\ref{fig:scc_comparison}), and the overall SCC levels become less sensitive to incremental load growth. This is because SCC is directly coupled with commitment status, and its magnitude changes only when load fluctuations are large enough to trigger the startup or shutdown of certain SGs, implying that newly committed SGs, which mainly provide SCC services, can accommodate increasing energy demand.

\subsubsection{Service Prices under Different Pricing Methods}\label{Service Prices under Different Pricing Methods}

\begin{figure}[!t]
\centering
\begin{tikzpicture}

\begin{axis}[
    width=0.97\linewidth, height=4cm,
    xlabel={\scriptsize Energy demand (GWh)},
    ylabel={\scriptsize Energy price (\texteuro/MWh)},
    label style={font=\scriptsize}, tick label style={font=\scriptsize},
    xmin=3.9, xmax=8.1,
    ymin=6, ymax=16,
    xtick={4.0,4.8,5.6,6.4,7.2,8.0},
    ytick={6,8,10,12,14,16},
    grid=both, grid style={dotted, gray!60},
    legend style={
        font=\scriptsize, 
        fill=none, 
        draw=none, 
        at={(0.17,0.75)}, anchor=north
    },
    legend cell align={left},
    legend image post style={scale=1},
    legend image code/.code={%
        \draw[mark repeat=1,mark phase=1,#1] plot coordinates {(0,0)};
    },
    axis y line*=left,
    axis x line*=bottom,
    axis lines=box
]

\addplot+[color=green!50!black, mark=o, line width=0.6pt]
    coordinates {
        (4.0,7.07)
        (4.8,7.2456)
        (5.6,8.72)
        (6.4,10.47)
        (7.2,12.3567)
        (8.0,15.02)
    };
\addlegendentry{$\lambda^\mathrm{E}$-Disp.}

\addplot+[color=blue, mark=o, line width=0.6pt]
    coordinates {
        (4.0,7.07)
        (4.8,7.10)
        (5.6,8.72)
        (6.4,8.72)
        (7.2,11.49)
        (8.0,15.02)
    };
\addlegendentry{$\lambda^\mathrm{E}$-Rest.}

\addplot+[color=red, mark=o, line width=0.6pt]
    coordinates {
        (4.0,7.10)
        (4.8,7.5777)
        (5.6,8.1012)
        (6.4,11.1309)
        (7.2,12.2614)
        (8.0,15.02)
    };
\addlegendentry{$\lambda^\mathrm{E}$-P-D}

\end{axis}

\begin{axis}[
    width=0.97\linewidth, height=4cm,
    xmin=3.9, xmax=8.1,
    ymin=15, ymax=42,
    axis y line*=right,
    axis x line=none,
    ylabel={\scriptsize SCC price (k\texteuro/p.u.)},
    xtick={4.0,4.8,5.6,6.4,7.2,8.0},
    ytick={20,25,30,35,40},
    tick label style={font=\scriptsize},
        tick align=outside, 
    ylabel style={at={(axis description cs:1.07,0.5)}},
    legend style={
        font=\scriptsize, 
        fill=none, 
        draw=none, 
        at={(0.86,1.04)}, anchor=north east,
        row sep=0pt,
    },
    legend cell align={left},
    legend image post style={scale=1},
    legend image code/.code={%
        \draw[mark repeat=1,mark phase=1,#1] plot coordinates {(0,0)};
    }
]

\addplot+[color=orange, mark=square*, line width=0.6pt]
    coordinates {
        (4.0,37.7003)
        (4.8,36.8103)
        (5.6,34.2491)
        (6.4,31.8658)
        (7.2,17.7594)
        (8.0,16.9783)
    };
\addlegendentry{$\lambda^\mathrm{SCC}_{26}$-Disp.}

\addplot+[color=purple, mark=square*, line width=0.6pt]
    coordinates {
        (4.0,40.4747)
        (4.8,38.0332)
        (5.6,37.7897)
        (6.4,30.9458)
        (7.2,19.5356)
        (8.0,17.5789)
    };
\addlegendentry{$\lambda^\mathrm{SCC}_{26}$-P-D}

\end{axis}

\end{tikzpicture}
\vspace{-0.8cm}
\caption{Prices of energy and SCC with different energy demand levels and pricing approaches. The SCC price for other buses is zero in each case.
}
\label{fig:price_scc_demand}
\vspace{-0.3cm}
\end{figure}

\begin{figure}[!t]
\centering
\begin{tikzpicture}

\begin{axis}[
    width=1.07\linewidth, height=4cm,
    xlabel={\scriptsize Energy demand (GWh)},
    ylabel={\scriptsize Commitment price (k\texteuro/h)},
    label style={font=\scriptsize}, tick label style={font=\scriptsize},
    xmin=3.9, xmax=8.1,
    ymin=-1, ymax=8,
    xtick={4.0,4.8,5.6,6.4,7.2,8.0},
    grid=both, grid style={dotted, gray!60},
        legend style={
        font=\scriptsize,
        fill=none,
        draw=none,
        at={(0.4,0.91)},
        anchor=north west,
        row sep=-2pt,
        legend columns=2,
    },
    legend cell align={left},
    legend image post style={scale=0.7},
    legend image code/.code={%
        \draw[mark repeat=1,mark phase=1,#1] plot coordinates {(0,0)};
    },
    axis y line*=left,
    axis x line*=bottom,
    axis lines=box
]

\addplot+[color=blue, solid, mark=o, line width=0.6pt] coordinates {(4.0,2.404) (4.8,2.395) (5.6,0) (6.4,1.905) (7.2,0.604) (8.0,0)}; \addlegendentry{$g_1$-$b4$}
\addplot+[color=red, solid, mark=o, line width=0.6pt] coordinates {(4.0,7.330) (4.8,7.326) (5.6,0) (6.4,0) (7.2,0) (8.0,0)}; \addlegendentry{$g_1$-$b27$}
\addplot+[color=green!50!black, thick, mark=o, line width=0.6pt] coordinates {(4.0,0) (4.8,0) (5.6,7.254) (6.4,7.254) (7.2,6.894) (8.0,6.217)}; \addlegendentry{$g_2$-$b27$}
\addplot+[color=orange, thick, mark=o, line width=0.6pt] coordinates {(4.0,4.415) (4.8,4.413) (5.6,4.320) (6.4,4.320) (7.2,4.160) (8.0,3.957)}; \addlegendentry{$g_2$-$b30$}

\end{axis}

\end{tikzpicture}
\vspace{-0.4cm}
\caption{Commitment price ($\lambda_{g,\mathrm{commit}}$) for SGs under different demand levels. For visual clarity, the commitment price for other SGs is not shown, as they exhibit similar trends to $g_1$-$b4$ and $g_1$-$b27$. While the price for $2g$-$b5$ is zero, since they are not dispatched.}
\label{fig:commitment_price}
     \vspace{-0.6cm}
\end{figure}

Fig.~\ref{fig:price_scc_demand} illustrates the impact of energy demand on the shadow prices under the three pricing methods. To satisfy growing demand, units with higher operating costs are gradually dispatched, driving up the energy price. Meanwhile, more SCC is provided to the system as a byproduct of additional committed thermal units for energy supply, thereby reducing the need for extra SCC procurement and lowering its price. Notably, only bus 26 procures SCC service in this period (as discussed in Section~\ref{SCC Improvement at Risk Buses}), whereas all other buses passively benefit from adequate SCC without additional procurement. Overall, in the presence of non-convexities, the single-period prices obtained from the P–D method do not deviate significantly from those of the dispatchable method, where integrality constraints are simply relaxed.

In addition, the restricted method fails to yield an explicit SCC price. This is because SCC is provided in a discrete manner, and the commitment variables are fixed at their optimal values, rendering the SCC constraint non-binding in the subsequent pricing stage. Instead, the resulting commitment prices are displayed in Fig.~\ref{fig:commitment_price}. As energy demand rises, the commitment prices of most SGs decline to zero, implying that these units no longer require make-whole payments, since their increasing energy revenues become sufficient to cover their costs. By contrast, units $g_2$-$b27$ and $g_2$-$b30$, which have relatively high operating costs, still rely on such payments to avoid losses.

\subsubsection{Profitability of Generators under Different Pricing Methods}\label{Profitability of Generators under Different Pricing Methods}
The units $2g$-$b5$ remain offline and are thus excluded from the analysis. Table~\ref{table:energy_profit} summarizes the energy profit of the units (energy revenue minus operating cost). The results show that, under each pricing method, multiple units are unable to fully recover their costs solely through the energy market. Therefore, additional remunerations such as SCC revenues (for dispatchable and P-D pricing) and make-whole payments (for restricted pricing) need to be appropriately allocated to the corresponding units; otherwise, they would not have sufficient incentives to follow the system operator’s dispatch and may exit the market.

The allocation of SCC revenues under the dispatchable method and the P‑D method is listed in Table~\ref{table:scc_revenue}. As indicated by the blue regions, since SCC prices under the dispatchable method are generally lower than those under the P‑D method (as shown in Fig.~\ref{fig:price_scc_demand}), the corresponding ancillary service revenue cannot always compensate for energy revenue shortfalls (the blue regions in Table~\ref{table:energy_profit}), leading to losses for units $g_2$-$b2$, $2g$-$b4$ and $2g$-$b30$ across various load levels. This indicates that, without appropriate make-whole payments, the service prices obtained from the dispatchable method would not support the dispatch. In contrast, the P‑D method optimizes shadow prices such that no unit incurs a loss without the need for any make-whole payment. 

These different market outcomes arise partly because the P‑D method explicitly enforces remuneration-sufficient shadow prices to accurately capture the contributions of thermal units. 
They also stem from the fact that the dispatchable method relaxes integrality constraints, allowing units to unrealistically connect only a portion of their impedances to the system during the pricing formulation. As a result, the SCC prices determined in this way only partially reflect the SCC provision of units, rather than fully capturing such contributions as the P-D method does. 

It is worth noting that, owing to the short electrical distance between units $2g$-$b27$ and bus 26 (the only SCC-procuring bus in this case), these units can provide substantial SCC and thus attain significant ancillary service revenues and total profits, as marked by the red regions in Tables~\ref{table:energy_profit} and \ref{table:scc_revenue}.

Although the commitment prices (Fig.~\ref{fig:commitment_price}) in the restricted method can fully offset operating losses of SGs (as shown in Fig.~\ref{fig:energy_profit_curve}), it remains unclear whether such payments remunerate units for being online to satisfy energy demand or merely to meet SCC requirements. Besides, the magnitude of these prices would depend on the energy profit shortfall of the units. For instance, the profit gap of $2g$-$b27$ (the gray regions in Table~\ref{table:energy_profit}) is covered across load levels, resulting in zero profit. This raises a concern: since $2g$-$b27$ are actually responsible for supplying major SCC at critical buses, they would not be incentivized to participate in the market if their final profit remains zero. It turns out that commitment prices would not intuitively reflect the economic value of SCC, and units still lack effective SCC price signals under this pricing method.

This core limitation naturally extends to synchronous condensers, which can likewise be modeled with binary on/off variables and can provide other ancillary services such as inertia. Under restricted pricing, the economic value of each `all or nothing' service is bundled together within the commitment decision. Consequently, explicit locational SCC shadow prices cannot be separated out, and the SCC contribution from synchronous condensers cannot be individually quantified. Therefore, restricted pricing is not suitable for targeted remuneration of their SCC service. This qualitative argument would hold generally for assets with binary commitment status which decides the provision of discrete services.

\subsubsection{Potential Lost Opportunity Costs}\label{Potential Uplift Payments for Lost Opportunity Costs}
Based on the definition in Section~\ref{Uplift Payment}, the LOC used here is:
\begin{equation}\label{eq:LOCs}
\vspace{-0.1cm}
\mathrm{LOC}_g \hspace{-0.05cm} = \hspace{-0.1cm} \max_{x_g\in X_g} \hspace{-0.1cm} \Pi_g(x_g;\lambda^\mathrm{E},\lambda_{b}^{\mathrm{SCC}})-\Pi_g(x_{g}^{\mathrm{SO}};\lambda^\mathrm{E},\lambda_{b}^{\mathrm{SCC}}) 
\end{equation}
where $X_g$ is the feasible operating set of SGs, and $x_g$ is a feasible operating point. Function $\Pi_g$ represents the generator’s profit evaluated at energy and SCC prices. The first term implies that, faced with posted prices, generators would seek to offer output $x_g$ that would maximize profits, while the second represents the profit at the system operator's dispatch $x_{g}^{\mathrm{SO}}$.

The LOCs of individual SGs at different load levels are shown in Fig.~\ref{fig:unit_uplift}. SGs at buses \{2, 3, 4\} have relatively low LOCs because they are generally dispatched close to their profit-maximizing points. When energy prices exceed marginal generation costs, these units would operate near their available capacities, leaving little room for further profit increases. In contrast, SGs at buses 27 and 30 exhibit substantially higher LOCs because energy prices are well below their marginal generation costs. At these prices, their actual dispatch yields considerably lower profits than those associated with their hypothetical profit-maximizing operating points, which may involve shutting down or operating at minimum stable generation. This difference is particularly pronounced when the dispatch requires these units to generate more power while providing SCC. As load levels and energy prices increase, the LOCs would decrease, indicating that the energy profit shortfalls under dispatch are better covered.

However, as clarified in Section~\ref{Relevant Economic Principles and Market Settings}, such LOCs are permissible within a quasi-partial equilibrium, and units would remain subject to dispatch. While these types of LOCs might be further reduced through CHP, that is a different topic, one that would require clear rules to determine who should bear these uplifts.

\begin{figure}[t]
\centering
\begin{tikzpicture}

\begin{axis}[
    width=1.05\linewidth, height=4cm,
    xlabel={\scriptsize Energy demand (GWh)},
    ylabel={\scriptsize Total profit (k\texteuro/h)},
    label style={font=\scriptsize}, tick label style={font=\scriptsize},
    xmin=3.9, xmax=8.1,
    ymin=-1, ymax=10,  
    xtick={4.0,4.8,5.6,6.4,7.2,8.0},
    grid=both, grid style={dotted, gray!60},
    legend style={
        font=\scriptsize,
        fill=none,
        draw=none,
        at={(0,0.93)},
        anchor=north west,
        row sep=-4pt,
        legend columns=4,
        /tikz/column 4/.style={column sep=0pt},
    },
    legend cell align={left},
    legend image post style={scale=0.7},
    legend image code/.code={%
        \draw[mark repeat=1,mark phase=1,#1] plot coordinates {(0,0)};
    },
    axis y line*=left,
    axis x line*=bottom,
    axis lines=box
]

\addplot+[color=blue, solid, mark=o, line width=0.6pt] coordinates {(4.0,0.00) (4.8,0.00) (5.6,1.58) (6.4,1.58) (7.2,5.22) (8.0,9.87)}; \addlegendentry{$g_1$-$b2$}
\addplot+[color=red, solid, mark=o, line width=0.6pt] coordinates {(4.0,0.00) (4.8,0.00) (5.6,0.43) (6.4,0.43) (7.2,4.08) (8.0,8.73)}; \addlegendentry{$g_2$-$b2$}
\addplot+[color=green, solid, mark=o, line width=0.6pt] coordinates {(4.0,0.00) (4.8,0.00) (5.6,0.37) (6.4,0.37) (7.2,3.56) (8.0,7.62)}; \addlegendentry{$g_1$-$b3$}
\addplot+[color=orange, solid, mark=o, line width=0.6pt] coordinates {(4.0,0.00) (4.8,0.00) (5.6,0.00) (6.4,0.00) (7.2,1.69) (8.0,5.76)}; \addlegendentry{$g_2$-$b3$}
\addplot+[color=purple, solid, mark=o, line width=0.6pt] coordinates {(4.0,0.00) (4.8,0.00) (5.6,0.00) (6.4,0.00) (7.2,0.00) (8.0,2.07)}; \addlegendentry{$g_1$-$b4$}
\addplot+[color=cyan, solid, mark=o, line width=0.6pt] coordinates {(4.0,0.00) (4.8,0.00) (5.6,0.00) (6.4,0.00) (7.2,0.00) (8.0,1.29)}; \addlegendentry{$g_2$-$b4$}

\end{axis}

\end{tikzpicture}
\vspace{-0.4cm}
\caption{Total profit (energy profit plus commitment price) for SGs under different demand levels, using restricted pricing. The profit for $2g$-$b5$ is zero since they are not dispatched. Meanwhile, the profit of the remaining SGs is also zero, because the commitment price exactly offsets their profit shortfall.}
\label{fig:energy_profit_curve}
\vspace{-0.2cm}
\end{figure}

\begin{figure}[t]
\centering
\begin{tikzpicture}
\begin{axis}[
    width=1\linewidth,
    height=4cm,
    xlabel={\scriptsize SGs' index},
    ylabel={\scriptsize Uplift (\texteuro/MWh)},
    label style={font=\scriptsize},
    tick label style={font=\scriptsize},
    xmin=0.5, xmax=12.5,
    xtick={1,...,12},
    xticklabels={
        $g_1$-$b2$, $g_2$-$b2$,
        $g_1$-$b3$, $g_2$-$b3$,
        $g_1$-$b4$, $g_2$-$b4$,
        $g_1$-$b5$, $g_2$-$b5$,
        $g_1$-$b27$, $g_2$-$b27$,
        $g_1$-$b30$, $g_2$-$b30$
    },
    x tick label style={
        rotate=45,
        anchor=east,
        font=\scriptsize
    },
    ymin=0,
    grid=both,
    grid style={dotted, gray!60},
    legend style={
        font=\scriptsize,
        fill=none,
        draw=none,
        at={(0,0.99)},
        anchor=north west,
        row sep=-4pt,
        legend columns=3,
        /tikz/column 3/.style={column sep=3pt}
    },
    legend cell align=left,
    legend image post style={scale=0.7},
    axis y line*=left,
    axis x line*=bottom,
    axis lines=box,
    clip=false,
    xlabel style={yshift=10pt}
]
\addplot+[blue, solid, mark=o, line width=0.6pt]
coordinates {
    (1, 0.228173) (2, 0.228165) (3, 0.220927) (4, 0.169369)
    (5, 0.234501) (6, 0.173243) (7, 0.0)      (8, 0.0)
    (9, 0.766129) (10, 6.05719) (11, 1.21371) (12, 1.26154)
};
\addlegendentry{4.0 GWh}

\addplot+[red, solid, mark=o, line width=0.6pt]
coordinates {
    (1, 0.178675) (2, 0.175705) (3, 0.201156) (4, 0.159363)
    (5, 0.183629) (6, 0.131785) (7, 0.0)      (8, 0.0)
    (9, 0.599929) (10, 4.6623)  (11, 0.950411) (12, 0.987868)
};
\addlegendentry{4.8 GWh}

\addplot+[green!60!black, solid, mark=o, line width=0.6pt]
coordinates {
    (1, 0.148631) (2, 0.146793) (3, 0.152177) (4, 0.187103)
    (5, 0.182087) (6, 0.13758)  (7, 0.0)      (8, 0.0)
    (9, 3.79407)  (10, 0.519116) (11, 0.809769) (12, 0.84065)
};
\addlegendentry{5.6 GWh}

\addplot+[orange, solid, mark=o, line width=0.6pt]
coordinates {
    (1, 0.156424) (2, 0.153755) (3, 0.163297) (4, 0.203305)
    (5, 0.222028) (6, 0.173815) (7, 0.0)      (8, 0.0)
    (9, 2.5765)   (10, 0.193231) (11, 0.483691) (12, 0.155628)
};
\addlegendentry{6.4 GWh}

\addplot+[purple, solid, mark=o, line width=0.6pt]
coordinates {
    (1, 0.087776) (2, 0.0862784) (3, 0.0916325) (4, 0.0971717)
    (5, 0.0982915) (6, 0.117634)  (7, 0.0)       (8, 0.0)
    (9, 1.11787)  (10, 0.10843)   (11, 0.0653484) (12, 0.0873293)
};
\addlegendentry{7.2 GWh}

\addplot+[cyan, solid, mark=o, line width=0.6pt]
coordinates {
    (1, 0.0710859) (2, 0.069873)  (3, 0.0742091) (4, 0.0786951)
    (5, 0.0796019) (6, 0.0789891) (7, 0.0)       (8, 0.0)
    (9, 0.963742)  (10, 0.0878127) (11, 0.0210081) (12, 0.0707241)
};
\addlegendentry{8.0 GWh}

\end{axis}
\end{tikzpicture}
\vspace{-0.3cm}
\caption{Lost opportunity costs of the SGs under different load levels.}
\label{fig:unit_uplift}
\vspace{-0.3cm}
\end{figure}

\begin{table}[!t]
\centering
\caption{Energy Profit of Thermal Units \\ under Different Conditions (k\texteuro)}
\vspace{-0.3cm}
\setlength{\tabcolsep}{3.5pt}
{\fontsize{7.5pt}{10pt}\selectfont
\begin{tabular}{l!{\vrule width 0.5pt}l@{\hspace{2pt}}cccccc}
\toprule
\multicolumn{2}{c}{Energy demand (GWh)} & 4.0 & 4.8 & 5.6 & 6.4 & 7.2 & 8.0 \\
\midrule
\multirow{12}{*}{\rotatebox{90}{\footnotesize \textcolor{blue}{Dispatchable pricing}}}
& $g_1$-$b2$   & $\mathrm{-0.60}$ & $\mathrm{-0.37}$ & $\mathrm{1.58}$ & $\mathrm{3.88}$ & $\mathrm{6.36}$ & $\mathrm{9.87}$ \\
& $g_2$-$b2$   & \cellcolor{blue!30}{$\mathrm{-1.74}$} & $\mathrm{-1.51}$ & $\mathrm{0.43}$ & $\mathrm{2.73}$ & $\mathrm{5.22}$ & $\mathrm{8.73}$ \\
& $g_1$-$b3$   & $\mathrm{-1.52}$ & $\mathrm{-1.37}$ & $\mathrm{0.37}$ & $\mathrm{2.38}$ & $\mathrm{4.56}$ & $\mathrm{7.62}$ \\
& $g_2$-$b3$   & $\mathrm{0.00}$ & $\mathrm{0.00}$ & $\mathrm{-1.50}$ & $\mathrm{0.38}$ & $\mathrm{2.69}$ & $\mathrm{5.76}$ \\
& $g_1$-$b4$   & \cellcolor{blue!30}{$\mathrm{-2.40}$} & \cellcolor{blue!30}{$\mathrm{-2.35}$} & $\mathrm{0.00}$ & $\mathrm{-1.38}$ & $\mathrm{0.05}$ & $\mathrm{2.07}$ \\
& $g_2$-$b4$   & $\mathrm{0.00}$ & $\mathrm{0.00}$ & $\mathrm{0.00}$ & \cellcolor{blue!30}{$\mathrm{-1.68}$} & \cellcolor{blue!30}{$\mathrm{-0.88}$} & $\mathrm{1.29}$ \\
& $g_1$-$b27$  & $\mathrm{-7.33}$ & $\mathrm{-7.31}$ & $\mathrm{0.00}$ & $\mathrm{0.00}$ & $\mathrm{0.00}$ & $\mathrm{0.00}$ \\
& $g_2$-$b27$  & $\mathrm{0.00}$ & $\mathrm{0.00}$ & $\mathrm{-7.25}$ & $\mathrm{-7.03}$ & $\mathrm{-6.78}$ & $\mathrm{-6.22}$ \\
& $g_1$-$b30$  & \cellcolor{blue!30}{$\mathrm{-4.43}$} & \cellcolor{blue!30}{$\mathrm{-4.42}$} & \cellcolor{blue!30}{$\mathrm{-4.34}$} & $\mathrm{0.00}$ & $\mathrm{0.00}$ & $\mathrm{0.00}$ \\
& $g_2$-$b30$  & $\mathrm{-4.41}$ & $\mathrm{-4.40}$ & $\mathrm{-4.32}$ & $\mathrm{-4.22}$ & \cellcolor{blue!30}{$\mathrm{-4.11}$} & \cellcolor{blue!30}{$\mathrm{-3.96}$} \\
\midrule
\multirow{12}{*}{\rotatebox{90}{\footnotesize Restricted pricing}}
& $g_1$-$b2$   & $\mathrm{-0.60}$ & $\mathrm{-0.56}$ & $\mathrm{1.58}$ & $\mathrm{1.58}$ & $\mathrm{5.22}$ & $\mathrm{9.87}$ \\
& $g_2$-$b2$   & $\mathrm{-1.74}$ & $\mathrm{-1.70}$ & $\mathrm{0.43}$ & $\mathrm{0.43}$ & $\mathrm{4.08}$ & $\mathrm{8.73}$ \\
& $g_1$-$b3$   & $\mathrm{-1.52}$ & $\mathrm{-1.50}$ & $\mathrm{0.37}$ & $\mathrm{0.37}$ & $\mathrm{3.56}$ & $\mathrm{7.62}$ \\
& $g_2$-$b3$   & $\mathrm{0.00}$ & $\mathrm{0.00}$ & $\mathrm{-1.50}$ & $\mathrm{-1.50}$ & $\mathrm{1.69}$ & $\mathrm{5.76}$ \\
& $g_1$-$b4$   & $\mathrm{-2.40}$ & $\mathrm{-2.39}$ & $\mathrm{0.00}$ & $\mathrm{-1.90}$ & $\mathrm{-0.60}$ & $\mathrm{2.07}$ \\
& $g_2$-$b4$   & $\mathrm{0.00}$ & $\mathrm{0.00}$ & $\mathrm{0.00}$ & $\mathrm{-2.21}$ & $\mathrm{-1.38}$ & $\mathrm{1.29}$ \\
& $g_1$-$b27$  & \cellcolor{gray!30}{$\mathrm{-7.33}$} & \cellcolor{gray!30}{$\mathrm{-7.33}$} & $\mathrm{0.00}$ & $\mathrm{0.00}$ & $\mathrm{0.00}$ & $\mathrm{0.00}$ \\
& $g_2$-$b27$  & $\mathrm{0.00}$ & $\mathrm{0.00}$ & \cellcolor{gray!30}{$\mathrm{-7.25}$} & \cellcolor{gray!30}{$\mathrm{-7.25}$} & \cellcolor{gray!30}{$\mathrm{-6.89}$} & \cellcolor{gray!30}{$\mathrm{-6.22}$} \\
& $g_1$-$b30$  & $\mathrm{-4.43}$ & $\mathrm{-4.43}$ & $\mathrm{-4.34}$ & $\mathrm{0.00}$ & $\mathrm{0.00}$ & $\mathrm{0.00}$ \\
& $g_2$-$b30$  & $\mathrm{-4.41}$ & $\mathrm{-4.41}$ & $\mathrm{-4.32}$ & $\mathrm{-4.32}$ & $\mathrm{-4.16}$ & $\mathrm{-3.96}$ \\
\midrule
\multirow{12}{*}{\rotatebox{90}{\footnotesize \textcolor{purple}{Primal-dual pricing} }}
& $g_1$-$b2$   & $\mathrm{-0.56}$ & $\mathrm{0.07}$ & $\mathrm{0.76}$ & $\mathrm{4.75}$ & $\mathrm{6.24}$ & $\mathrm{9.87}$ \\
& $g_2$-$b2$   & $\mathrm{-1.72}$ & $\mathrm{-1.07}$ & $\mathrm{-0.39}$ & $\mathrm{3.60}$ & $\mathrm{5.09}$ & $\mathrm{8.73}$ \\
& $g_1$-$b3$   & $\mathrm{-1.50}$ & $\mathrm{-1.09}$ & $\mathrm{-0.35}$ & $\mathrm{3.14}$ & $\mathrm{4.45}$ & $\mathrm{7.62}$ \\
& $g_2$-$b3$   & $\mathrm{0.00}$ & $\mathrm{0.00}$ & $\mathrm{-2.01}$ & $\mathrm{1.08}$ & $\mathrm{2.58}$ & $\mathrm{5.76}$ \\
& $g_1$-$b4$   & $\mathrm{-2.39}$ & $\mathrm{-2.25}$ & $\mathrm{0.00}$ & $\mathrm{-1.18}$ & $\mathrm{-0.02}$ & $\mathrm{2.07}$ \\
& $g_2$-$b4$   & $\mathrm{0.00}$ & $\mathrm{0.00}$ & $\mathrm{0.00}$ & $\mathrm{-1.48}$ & $\mathrm{-0.94}$ & $\mathrm{1.29}$ \\
& $g_1$-$b27$  & \cellcolor{red!30}{$\mathrm{-7.33}$} & \cellcolor{red!30}{$\mathrm{-7.26}$} & $\mathrm{0.00}$ & $\mathrm{0.00}$ & $\mathrm{0.00}$ & $\mathrm{0.00}$ \\
& $g_2$-$b27$  & $\mathrm{0.00}$ & $\mathrm{0.00}$ & \cellcolor{red!30}{$\mathrm{-7.33}$} & \cellcolor{red!30}{$\mathrm{-6.94}$} & \cellcolor{red!30}{$\mathrm{-6.79}$} & \cellcolor{red!30}{$\mathrm{-6.22}$} \\
& $g_1$-$b30$  & $\mathrm{-4.43}$ & $\mathrm{-4.41}$ & $\mathrm{-4.38}$ & $\mathrm{0.00}$ & $\mathrm{0.00}$ & $\mathrm{0.00}$ \\
& $g_2$-$b30$  & $\mathrm{-4.41}$ & $\mathrm{-4.39}$ & $\mathrm{-4.36}$ & $\mathrm{-4.18}$ & $\mathrm{-4.12}$ & $\mathrm{-3.96}$ \\
\bottomrule
\end{tabular}
}
\label{table:energy_profit}
\vspace{-0.4cm}
\end{table}

\begin{table}[!t]
\centering
\caption{SCC Revenue of Thermal Units \\ under Different Conditions (k\texteuro)}
\vspace{-0.3cm}
\setlength{\tabcolsep}{4.7pt}
{\fontsize{7.5pt}{10pt}\selectfont
\begin{tabular}{l!{\vrule width 0.5pt}l@{\hspace{2pt}}cccccc}
\toprule
\multicolumn{2}{c}{Energy demand (GWh)} & 4.0 & 4.8 & 5.6 & 6.4 & 7.2 & 8.0 \\
\midrule
\multirow{12}{*}{\rotatebox{90}{\footnotesize \textcolor{blue}{Dispatchable pricing} }}
& $g_1$-$b2$   & $\mathrm{1.65}$ & $\mathrm{1.61}$ & $\mathrm{1.52}$ & $\mathrm{1.08}$ & $\mathrm{0.60}$ & $\mathrm{0.58}$ \\
& $g_2$-$b2$   & \cellcolor{blue!30}{$\mathrm{1.73}$} & $\mathrm{1.69}$ & $\mathrm{1.58}$ & $\mathrm{1.15}$ & $\mathrm{0.64}$ & $\mathrm{0.61}$ \\
& $g_1$-$b3$   & $\mathrm{1.96}$ & $\mathrm{1.91}$ & $\mathrm{1.75}$ & $\mathrm{1.28}$ & $\mathrm{0.71}$ & $\mathrm{0.68}$ \\
& $g_2$-$b3$   & $\mathrm{0.00}$ & $\mathrm{0.00}$ & $\mathrm{1.82}$ & $\mathrm{1.31}$ & $\mathrm{0.73}$ & $\mathrm{0.70}$ \\
& $g_1$-$b4$   & \cellcolor{blue!30}{$\mathrm{2.23}$} & \cellcolor{blue!30}{$\mathrm{2.18}$} & $\mathrm{0.00}$ & $\mathrm{1.47}$ & $\mathrm{0.82}$ & $\mathrm{0.78}$ \\
& $g_2$-$b4$   & $\mathrm{0.00}$ & $\mathrm{0.00}$ & $\mathrm{0.00}$ & \cellcolor{blue!30}{$\mathrm{1.53}$} & \cellcolor{blue!30}{$\mathrm{0.85}$} & $\mathrm{0.81}$ \\
& $g_1$-$b27$  & $\mathrm{25.52}$ & $\mathrm{24.92}$ & $\mathrm{0.00}$ & $\mathrm{0.00}$ & $\mathrm{0.00}$ & $\mathrm{0.00}$ \\
& $g_2$-$b27$  & $\mathrm{0.00}$ & $\mathrm{0.00}$ & $\mathrm{24.18}$ & $\mathrm{23.68}$ & $\mathrm{13.20}$ & $\mathrm{12.62}$ \\
& $g_1$-$b30$  & \cellcolor{blue!30}{$\mathrm{4.37}$} & \cellcolor{blue!30}{$\mathrm{4.26}$} & \cellcolor{blue!30}{$\mathrm{3.97}$} & $\mathrm{0.00}$ & $\mathrm{0.00}$ & $\mathrm{0.00}$ \\
& $g_2$-$b30$  & $\mathrm{5.00}$ & $\mathrm{4.88}$ & $\mathrm{4.55}$ & $\mathrm{7.17}$ & \cellcolor{blue!30}{$\mathrm{4.00}$} & \cellcolor{blue!30}{$\mathrm{3.82}$} \\
\midrule
\multirow{12}{*}{\rotatebox{90}{\footnotesize \textcolor{purple}{Primal-dual pricing} }}
& $g_1$-$b2$   & $\mathrm{1.77}$ & $\mathrm{1.67}$ & $\mathrm{1.67}$ & $\mathrm{1.05}$ & $\mathrm{0.66}$ & $\mathrm{0.60}$ \\
& $g_2$-$b2$   & $\mathrm{1.85}$ & $\mathrm{1.74}$ & $\mathrm{1.75}$ & $\mathrm{1.12}$ & $\mathrm{0.71}$ & $\mathrm{0.64}$ \\
& $g_1$-$b3$   & $\mathrm{2.10}$ & $\mathrm{1.98}$ & $\mathrm{1.94}$ & $\mathrm{1.24}$ & $\mathrm{0.78}$ & $\mathrm{0.70}$ \\
& $g_2$-$b3$   & $\mathrm{0.00}$ & $\mathrm{0.00}$ & $\mathrm{2.01}$ & $\mathrm{1.27}$ & $\mathrm{0.80}$ & $\mathrm{0.72}$ \\
& $g_1$-$b4$   & $\mathrm{2.39}$ & $\mathrm{2.25}$ & $\mathrm{0.00}$ & $\mathrm{1.43}$ & $\mathrm{0.90}$ & $\mathrm{0.81}$ \\
& $g_2$-$b4$   & $\mathrm{0.00}$ & $\mathrm{0.00}$ & $\mathrm{0.00}$ & $\mathrm{1.48}$ & $\mathrm{0.94}$ & $\mathrm{0.84}$ \\
& $g_1$-$b27$  & \cellcolor{red!30}{$\mathrm{27.40}$} & \cellcolor{red!30}{$\mathrm{25.75}$} & $\mathrm{0.00}$ & $\mathrm{0.00}$ & $\mathrm{0.00}$ & $\mathrm{0.00}$ \\
& $g_2$-$b27$  & $\mathrm{0.00}$ & $\mathrm{0.00}$ & \cellcolor{red!30}{$\mathrm{26.68}$} & \cellcolor{red!30}{$\mathrm{23.00}$} & \cellcolor{red!30}{$\mathrm{14.52}$} & \cellcolor{red!30}{$\mathrm{13.06}$} \\
& $g_1$-$b30$  & $\mathrm{4.69}$ & $\mathrm{4.41}$ & $\mathrm{4.38}$ & $\mathrm{0.00}$ & $\mathrm{0.00}$ & $\mathrm{0.00}$ \\
& $g_2$-$b30$  & $\mathrm{5.37}$ & $\mathrm{5.04}$ & $\mathrm{5.02}$ & $\mathrm{6.97}$ & $\mathrm{4.40}$ & $\mathrm{3.96}$ \\
\bottomrule
\end{tabular}
}
\label{table:scc_revenue}
\vspace{-0.3cm}
\end{table}

\subsection{Multi-Period SCC Service Pricing and Market Outcomes}\label{Multi-Period SCC Service Pricing}
\begin{table}[t]
\centering
\caption{Solution Performance of Primal-Dual Formulation under Multi-Period Operating Conditions}
     \vspace{-0.3cm}
\setlength{\tabcolsep}{8pt}
{\fontsize{8pt}{12pt}\selectfont
\begin{tabular}{l!{\vrule width 0.5pt}cc}
\toprule
\multicolumn{1}{c}{} & Primal obj. (M\texteuro) & Dual obj. (M\texteuro) \\
\midrule
\multirow{1}{*}{w/o \eqref{eq:nonnegative_profits_cons}} 
& $\mathrm{1.301}$ & $\mathrm{1.278}$ \\ 
\midrule
\multirow{1}{*}{with \eqref{eq:nonnegative_profits_cons}, $\forall t$} 
& $\mathrm{1.308}$  & $\mathrm{1.250}$ \\ 
\bottomrule
\end{tabular}
}
\label{table:solving_time_gap_multiperiod}
     \vspace{-0.1cm}
\end{table}

In the multi-period analysis, eqs.~\eqref{eq:multiperiod_PD} need to be incorporated to form the 24-hour market clearing. The energy demand within this market horizon fluctuates between 5.13 GWh and 7.69 GWh. Since Section~\ref{Profitability of Generators under Different Pricing Methods} has demonstrated that the restricted method is unsuitable for deriving needed SCC service prices, only the dispatchable method and the P‑D method are analyzed here. Note that \eqref{eq:nonnegative_profits_cons} is imposed hourly here for each SG to ensure no generator has incentives to exit the market in any dispatch interval. Analysis of the formulation that guarantees non-negative total profit over the entire horizon is presented in Section~\ref{Horizon-Aggregated Revenue Adequacy}.

The solution of the P-D method for full-day market clearing is presented in Table~\ref{table:solving_time_gap_multiperiod}. Similar to the single-period case, after introducing non-negative profit constraints, the dual objective value decreases by $2.19\%$ (from $1.278$ M\texteuro~to $1.250$ M\texteuro) to obtain the desired service prices, while the primal objective value only deviates from the optimal value by $0.54\%$ (from $1.301$ M\texteuro~to $1.308$ M\texteuro).

\subsubsection{SCC Improvement at Risky Buses}\label{SCC Improvement at Risky Buses_multiperiod}

\begin{figure}[!t]
\centering
\begin{tikzpicture}
\begin{axis}[
    width=1.05\linewidth, height=4cm,
    xlabel={\scriptsize Bus},
    ylabel={\scriptsize SCC (p.u.)}, 
    ylabel style={align=center, yshift=-0.1cm},
    label style={font=\scriptsize}, 
    tick label style={font=\tiny},
    xmin=1, xmax=30,
    ymin=0, ymax=15,
    xtick={1,2,...,30},
    grid=both, grid style={dotted, gray!60},
    legend style={
        font=\scriptsize, 
        fill=none, 
        draw=none, 
        at={(0.132,1.0)}, anchor=north west,
        legend columns=2,
        column sep=0pt,
    },
    legend cell align={left},
    axis lines=box,
    ybar,
    bar shift=0pt,
    bar width=4pt,
    enlarge x limits=0.02,
    legend image code/.code={%
        \draw[#1] (0,-0.1cm) rectangle (0.2cm,0.1cm);%
    },
]
\addplot[blue!20, fill=blue!20, draw=blue!70] coordinates {
(1,10.0269)  (2,10.2449)  (3,10.5452)  (4,10.5425)  (5,9.6032)   (6,11.3345)  (7,9.8258)   (8,10.4883)  (9,9.3366)   (10,9.6171)
(11,6.2747)  (12,9.1282)  (13,7.0138)  (14,7.4968)  (15,8.6697)  (16,7.7000)  (17,8.9538)  (18,7.4665)  (19,7.4729)  (20,7.5091)
(21,9.2425)  (22,9.1519)  (23,7.8102)  (24,8.8679)  (25,9.2868)  (26,5.0246)  (27,12.3612) (28,10.5961) (29,8.4501)  (30,8.8518)
};
\addlegendentry{with SCC constraints}

\addplot[red!20, fill=red!20, draw=red!70, opacity=0.8] coordinates {
(1,7.8005)   (2,7.5571)   (3,7.5103)   (4,7.5204)   (5,6.7865)   (6,7.9505)   (7,6.9784)   (8,7.0790)   (9,6.7010)   (10,6.5915)
(11,4.9366)  (12,6.3966)  (13,5.2485)  (14,5.6183)  (15,6.2300)  (16,5.7126)  (17,6.2167)  (18,5.5507)  (19,5.5576)  (20,5.6560)
(21,6.3076)  (22,6.3200)  (23,5.7820)  (24,5.9393)  (25,5.4781)  (26,3.9230)  (27,5.4976)  (28,7.0450)  (29,3.5690)  (30,2.6483)
};
\addlegendentry{w/o SCC constraints}

\end{axis}
\end{tikzpicture}
\vspace{-0.8cm}
\caption{Minimum SCC level at each bus with/without SCC constraints over the market horizon. The SCC threshold $\mathrm{I}_{b_{\mathrm{lim}}} = \mathrm{5~p.u.}$}
\label{fig:scc_bar_overlap}
     \vspace{-0.6cm}
\end{figure}

The distribution of minimum SCC, i.e., min$\{I_{b_\mathrm{L},t}~|~ \forall t\}$, at each system bus is depicted in Fig.~\ref{fig:scc_bar_overlap}. Evidently, buses \{11, 26, 29, 30\} fail to satisfy the protection device requirements over the entire time horizon. Buses \{11, 29\} are not connected to any local generation units, and thus rely entirely on SCC support from other buses. Moreover, their large electrical distance from the cheapest SGs (which are normally online for energy and SCC supply) restricts adequate SCC inflow (high-cost units in neighboring buses 27 and 30 are rarely dispatched). Buses \{26, 30\} also show a lack of local SCC support: bus 26 is equipped with only one wind turbine that provides very limited current injection, while bus 30 includes two usually offline SGs due to their high costs. 

Once SCC constraints are integrated into the UC, economic signals are generated to financially incentivize generator operation, ensuring that the system-wide SCC level stays above the security threshold across all time periods. 

\subsubsection{Service Prices under Different Pricing Methods}\label{Service Prices under Different Pricing Methods_multiperiod}
The SCC and energy prices calculated by the P-D and dispatchable methods are presented in Fig.~\ref{fig:price_split}. In this operation cycle, once bus 26 is secured with the required SCC level, other buses can passively receive the resultant SCC without requiring any further contribution, since only bus 26 yields a non-zero SCC price under the P-D method. Multi-period tests on SCC-constrained UC \eqref{eq:primal_model} also demonstrate that solely securing bus 26 is sufficient to maintain system SCC within the safe range. However, the dispatchable method generates redundant SCC price signals at unnecessary buses, e.g., bus 30 (a problem not observed in the single-period case as in Fig.~\ref{fig:price_scc_demand}). Furthermore, the SCC service price for the most critical bus, i.e., bus 26, is underestimated. These results reveal that relaxing UC integrality and neglecting the term `$\eta_{m,t}$' in the dispatchable method would lead to inefficient SCC prices. This highlights the necessity of preserving the binary nature to capture realistic system operating points and achieve accurate SCC security expression during the price formation.

It is also interesting to note that under the P-D pricing scheme, SCC and energy prices exhibit a complementary trend, especially at 01:00 and 10:00. Specifically, as several offline units are committed at the start of the scheduling horizon and incur startup costs, the SCC price at 01:00 is markedly higher than other periods. This permits lower energy prices while guaranteeing non-negative profits for thermal units. At 10:00, no SCC service is required and its price drops to zero; accordingly, the energy price surges to maintain financial wholeness of committed units. 

Such relationship arises because SCC is a byproduct when SGs are primarily dispatched for energy supply. Conversely, SGs have to sustain minimum stable generation during operation for securing SCC levels. In contrast, the dispatchable method fails to clearly capture this inherent coupling. It produces an energy price spike at 18:00 merely to match peak power demand instead of securing sufficient profits for relevant units. This demonstrates that relaxing integrality constraints weakens the intrinsic interdependence between energy and SCC markets.

\begin{figure}[!t]
\centering

\begin{minipage}[b]{1.03\linewidth}
\centering
\begin{tikzpicture}[baseline]
\begin{axis}[
    width=\linewidth, height=4cm,
    xlabel={\scriptsize Hour},
    ylabel={\scriptsize SCC price (k\texteuro/p.u.)},
    label style={font=\scriptsize}, tick label style={font=\scriptsize},
    xmin=1, xmax=24,
    ymin=0, ymax=36,
    xtick={1,2,...,24},
    grid=both, grid style={dotted, gray!60},
    legend style={
        font=\scriptsize, fill=none, draw=none,
        at={(0.3,1.0)}, anchor=north
    },
    legend cell align={left},
    axis y line*=left, axis x line*=bottom, axis lines=box
]

\addplot+[color=purple, mark=o, line width=0.6pt] coordinates {
(1,34.1044)(2,6.8382)(3,6.7981)(4,6.7831)(5,6.8236)(6,6.8460)(7,6.6578)(8,7.7827)
(9,4.9939)(10,0.0000)(11,4.7978)(12,5.6262)(13,5.6672)(14,5.5995)(15,4.7143)(16,5.6536)
(17,4.4061)(18,4.4061)(19,4.7143)(20,4.8935)(21,7.2890)(22,6.8483)(23,10.6851)(24,10.1809)
}; \addlegendentry{$\lambda^\mathrm{SCC}_{26}$-P-D};

\addplot+[color=orange, mark=o, line width=0.6pt] coordinates {
(1,8.8443)(2,2.2917)(3,2.2917)(4,2.0154)(5,2.2917)(6,2.2917)(7,2.0154)(8,1.4404)
(9,1.4404)(10,1.1274)(11,1.4118)(12,0.0550)(13,0.0000)(14,0.0000)(15,0.0000)(16,0.0000)
(17,0.0000)(18,0.0000)(19,0.0000)(20,0.0000)(21,0.0000)(22,1.6580)(23,1.8180)(24,8.6842)
}; \addlegendentry{$\lambda^\mathrm{SCC}_{26}$-Disp.};

\addplot+[color=cyan!70!blue, mark=o, line width=0.6pt] coordinates {
(1,7.194)(2,2.525)(3,2.525)(4,2.447)(5,2.525)(6,2.525)(7,2.447)(8,2.284)
(9,2.284)(10,2.275)(11,0.0000)(12,3.639)(13,3.655)(14,3.357)(15,0.0000)(16,0.0000)
(17,0.0000)(18,0.0000)(19,0.0000)(20,0.0000)(21,0.0000)(22,1.878)(23,2.391)(24,7.149)
}; \addlegendentry{$\lambda^\mathrm{SCC}_{30}$-Disp. ($\times10$)};

\end{axis}
\end{tikzpicture}
\end{minipage}
\hfill
\begin{minipage}[b]{1.03\linewidth}
\centering
\begin{tikzpicture}[baseline]
\begin{axis}[
    width=\linewidth, height=4cm,
    xlabel={\scriptsize Hour},
    ylabel={\scriptsize Energy price (\texteuro/MWh)},
    label style={font=\scriptsize}, tick label style={font=\scriptsize},
    xmin=1, xmax=24,
    ymin=5, ymax=26,
    xtick={1,2,...,24},
    grid=both, grid style={dotted, gray!60},
    legend style={
        font=\scriptsize, fill=none, draw=none,
        at={(0.2,1.0)}, anchor=north
    },
    legend cell align={left},
    axis y line*=left, axis x line*=bottom, axis lines=box
]

\addplot+[color=red, mark=o, line width=0.6pt] coordinates {
(1,7.8347)  (2,9.9935)  (3,10.0589) (4,10.0833) (5,10.0173) (6,9.9807)  (7,10.2877) (8,10.4700)
(9,13.4243) (10,24.5212)(11,13.2399)(12,11.8445)(13,11.8414)(14,11.9832)(15,13.5300)(16,11.8423)
(17,14.6000)(18,14.6000)(19,13.5300)(20,12.9546)(21,11.7131)(22,9.9771) (23,9.4757) (24,9.5015)
}; \addlegendentry{$\lambda^\mathrm{E}$-P-D};

\addplot+[color=green!50!black, mark=o, line width=0.6pt] coordinates {
(1,8.4686)  (2,7.1000)  (3,7.1000)  (4,8.7200)  (5,7.1000)  (6,7.1000)  (7,8.7200)  (8,12.0920)
(9,12.0920)(10,13.5300)(11,13.5300)(12,13.5256)(13,13.5300)(14,13.5300)(15,13.5300)(16,13.5300)
(17,14.6000)(18,21.8898)(19,13.5300)(20,13.3090)(21,12.2890)(22,9.8960)(23,9.8779)(24,9.4077)
}; \addlegendentry{$\lambda^\mathrm{E}$-Disp.};

\end{axis}
\end{tikzpicture}
\end{minipage}
\vspace{-0.4cm}
\caption{Price profiles of SCC (upper) and energy (lower) over the complete market horizon. The SCC price for other buses is zero in each case.}
\label{fig:price_split}
     \vspace{-0.2cm}
\end{figure}

\subsubsection{Profitability of Generators under Different Pricing Methods}\label{Profitability of Generators under Different Pricing Methods_multiperiod}

\begin{figure}[!t]
\centering
\begin{tikzpicture}
\begin{axis}[
    width=1.05\linewidth, height=4cm,
    xlabel={\scriptsize SGs' index},
    ylabel={\scriptsize Revenue and profit (k\texteuro)}, 
    ylabel style={align=center, yshift=-0.1cm, xshift=-6pt},
    label style={font=\scriptsize}, 
    tick label style={font=\scriptsize},
    xmin=1, xmax=12,
    ymin=-50, ymax=160,
    xtick={1,2,...,12},
    xticklabels={$g_1$-$b2$,$g_2$-$b2$,$g_1$-$b3$,$g_2$-$b3$,$g_1$-$b4$,$g_2$-$b4$,$g_1$-$b5$,$g_2$-$b5$,$g_1$-$b27$,$g_2$-$b27$,$g_1$-$b30$,$g_2$-$b30$},
       x tick label style={
        rotate=45,
        anchor=east,
        font=\scriptsize
    },
    grid=both, grid style={dotted, gray!60},
    legend style={
        font=\tiny, 
        fill=none, 
        draw=none, 
        at={(0.35,0.3)}, anchor=south west,
        legend columns=1,
        row sep=-2pt,
    },
    legend cell align={left},
    axis lines=box,
    enlarge x limits=0.04,
        xlabel style={yshift=10pt}
]

\addplot[
ybar,
area legend,
bar width=4pt,
fill=red!30,
draw=red!80
] coordinates {
(1,149.2047) (2,121.7878) (3,101.9810) (4,62.3725)
(5,8.8725) (6,0) (7,0) (8,0)
(9,86.6281) (10,86.1286) (11,0) (12,16.2658)
};
\addlegendentry{Total profit-P-D}

\addplot[
ybar,
area legend,
bar width=4pt,
fill=blue!20,
draw=blue!80,
] coordinates {
(1,131.5743) (2,104.2830) (3,90.8735) (4,48.6948)
(5,8.5369) (6,0) (7,0) (8,0)
(9,-19.3112) (10,-23.4463) (11,0) (12,-4.6014)
};
\addlegendentry{Total profit-Disp.}

\addplot[
red!80,
dashed,
mark=o,
mark size=1.4pt,
] coordinates {
(1,140.5486) (2,112.9109) (3,91.2598) (4,52.5295)
(5,5.3510) (6,0) (7,0) (8,0)
(9,-34.7620) (10,-40.1379) (11,0) (12,-28.0829)
};
\addlegendentry{Energy profit-P-D}

\addplot[
blue!80,
dashed,
mark=diamond,
mark size=1.4pt,
] coordinates {
(1,128.2929) (2,100.7967) (3,87.1101) (4,45.7989)
(5,8.0335) (6,0) (7,0) (8,0)
(9,-32.7343) (10,-38.0451) (11,0) (12,-28.8654)
};
\addlegendentry{Energy profit-Disp.}

\addplot[
red!70!black,
mark=o,
mark size=1.4pt,
] coordinates {
(1,8.6560) (2,8.8769) (3,10.7211) (4,9.8430)
(5,3.5214) (6,0) (7,0) (8,0)
(9,121.3901) (10,126.2665) (11,0) (12,44.3487)
};
\addlegendentry{SCC revenue-P-D}

\addplot[
blue!70!black,
mark=diamond,
mark size=1.4pt,
] coordinates {
(1,3.2814) (2,3.4863) (3,3.7634) (4,2.8959)
(5,0.5034) (6,0) (7,0) (8,0)
(9,13.4231) (10,14.5988) (11,0) (12,24.2640)
};
\addlegendentry{SCC revenue-Disp.}

\end{axis}
\end{tikzpicture}
\vspace{-0.8cm}
\caption{Profitability of each SG under P-D and dispatchable pricing methods. Total profit is equal to the sum of energy profit and SCC revenue, in which the energy profit is energy revenue minus operating cost.}
\label{fig:profit_12units}
     \vspace{-0.4cm}
\end{figure}

Fig.~\ref{fig:profit_12units} illustrates the profitability of each SG. It can be seen that units $2g$-$b4$, $2g$-$b5$ and $g_1$-$b30$ are rarely dispatched and thus do not achieve significant profits, while units with lower operating costs ($2g$-$b2$ and $2g$-$b3$) gain substantial profits from the energy market. However, owing to their long electrical distance from bus 26, where ancillary services are procured, they provide only limited SCC and consequently earn little SCC revenue. 

For all aforementioned units, both the P-D and dispatchable methods yield suitable prices to keep them financially whole. Nevertheless, for high-cost units $2g$-$b27$, which provide indispensable SCC support to bus 26 (as discussed in Section~\ref{Profitability of Generators under Different Pricing Methods}), SCC prices from the dispatchable method remain inadequate to offset their energy profit shortfalls, resulting in negative total profits. This is very likely to hinder these units from participating in the market and to compromise SCC security; moreover, higher system operating costs may be incurred when alternative SGs are inefficiently committed to compensate for the loss of critical SCC volumes. The P-D method, however, is able to capture the value of such critical SCC provision and allocate sufficient SCC remuneration to $2g$-$b27$. A similar behavior is observed for unit $g_2$-$b30$.

This comparison further reinforces the necessity of retaining integrality constraints to form efficient price signals that can adequately remunerate units. It also suggests that the cost characteristics and electrical locations of generators endow them with distinct operational roles. Accordingly, individual units should receive appropriate remuneration in respective markets to sustain economical and stable system operation. In this case, for instance, $2g$-$b2$ and $2g$-$b3$ are mainly compensated via the energy market, whereas $2g$-$b27$ obtain their remuneration primarily from the SCC market.

Note that across all test cases, only the activation status of the non-negative profit constraint varied, while all other model parameters and constraints remained unchanged. Consequently, the revenue difference between the dispatchable method and the P-D method shown in Fig.~\ref{fig:profit_12units} stems entirely from this revenue-adequacy requirement. Although the binary expansion adopted to linearize the bilinear price-quantity product introduces approximation error, its impact is not significant (as discussed in Section~\ref{Sensitivity of Pricing to Binary Expansion Errors}) and does not intentionally inflate the profit of generating units.

\subsubsection{Market Outcomes with Horizon-Aggregated Non-Negative Profits}\label{Horizon-Aggregated Revenue Adequacy}

\begin{figure}[!t]
\centering

\begin{tikzpicture}
\begin{axis}[
    width=0.95\linewidth,
    height=4.0cm,
    xlabel={\scriptsize Hour},
    ylabel={\scriptsize Energy price (\texteuro/MWh)},
    label style={font=\scriptsize},
    tick label style={font=\scriptsize},
    xmin=1, xmax=24,
    ymin=0, ymax=30,
    xtick={1,2,...,24},
    grid=both,
    grid style={dotted, gray!60},
    axis lines=box,
    axis y line*=left,
    axis x line*=bottom,
    legend style={
        font=\scriptsize,
        fill=none,
        draw=none,
        at={(0.02,1)},
        anchor=north west,
    },
    legend cell align=left,
    clip=false,
]

\addplot+[
    color=blue,
    mark=o,
    mark size=1.8pt,
    line width=0.6pt
] coordinates {
    (1, 7.1)
    (2, 7.1)
    (3, 7.1)
    (4, 7.311560634542212)
    (5, 7.1)
    (6, 7.1)
    (7, 8.72)
    (8, 11.878960276450432)
    (9, 11.970878006115557)
    (10, 13.53)
    (11, 13.53)
    (12, 13.53)
    (13, 13.53)
    (14, 13.53)
    (15, 13.53)
    (16, 13.53)
    (17, 15.02)
    (18, 15.36)
    (19, 14.6)
    (20, 13.53)
    (21, 11.87896027645043)
    (22, 9.720265381758834)
    (23, 9.162494120041845)
    (24, 8.72)
};
\addlegendentry{$\lambda^\mathrm{E}$};

\end{axis}

\begin{axis}[
    width=0.95\linewidth,
    height=4.0cm,
    xmin=1, xmax=24,
    ymin=0, ymax=40,
    xtick={1,2,...,24},
    ytick={0,10,20,30,40},
    axis y line*=right, 
    ylabel={\scriptsize SCC price (k\texteuro/p.u.)},
    label style={font=\scriptsize},
    tick label style={font=\scriptsize},
    ylabel near ticks,
    grid=none,
    legend style={
        font=\scriptsize,
        fill=none,
        draw=none,
        at={(0.02,0.90)},
        anchor=north west,
    },
    legend cell align=left,
    clip=false,
]

\addplot+[
    color=purple,
    mark=o,
    mark size=1.8pt,
    line width=0.6pt
] coordinates {
    (1, 34.219842400638845)
    (2, 0.0)
    (3, 0.0)
    (4, 21.14354007062244)
    (5, 0.0)
    (6, 0.0)
    (7, 19.064321234316714)
    (8, 5.189792034717656)
    (9, 5.163318848168487)
    (10, 0.0)
    (11, 1.295725179096236)
    (12, 11.822402513423891)
    (13, 4.0938941146197876)
    (14, 0.0)
    (15, 7.163217451296846)
    (16, 0.0)
    (17, 9.835618475062738)
    (18, 0.0)
    (19, 0.0)
    (20, 0.0)
    (21, 5.189792034717657)
    (22, 7.446152903381448)
    (23, 8.011198378005707)
    (24, 34.24908218554254)
};
\addlegendentry{$\lambda^\mathrm{SCC}_{26}$};

\end{axis}
\end{tikzpicture}


\begin{tikzpicture}
\begin{axis}[
    width=1\linewidth,
    height=4.0cm,
    xlabel={\scriptsize SGs' index},
    ylabel={\scriptsize Total profit (k\texteuro)},
    label style={font=\scriptsize},
    tick label style={font=\scriptsize},
    xmin=1, xmax=12,
    ymin=0, ymax=160,
    xtick={1,2,...,12},
    xticklabels={
        {$g_1$-$b2$},
        {$g_2$-$b2$},
        {$g_1$-$b3$},
        {$g_2$-$b3$},
        {$g_1$-$b4$},
        {$g_2$-$b4$},
        {$g_1$-$b5$},
        {$g_2$-$b5$},
        {$g_1$-$b27$},
        {$g_2$-$b27$},
        {$g_1$-$b30$},
        {$g_2$-$b30$}
    },
       x tick label style={
        rotate=45,
        anchor=east,
        font=\scriptsize
    },
    ymajorgrids=true,
    xmajorgrids=false,
    grid style={dotted, gray!60},
    axis lines=box,
    y tick label style={font=\scriptsize},
    legend style={
        font=\scriptsize,
        fill=none,
        draw=none,
        at={(0.15,0.98)},
        anchor=north west,
        legend columns=1,
    },
    legend cell align=left,
        xlabel style={yshift=10pt}
]

\addplot+[
    mark=o,
    color=blue,
    line width=0.6pt,
] coordinates {
    (1, 123.82935789587766)
    (2, 96.82796601389477)
    (3, 85.92458443200661)
    (4, 46.16182194497186)
    (5, 6.292409104689907)
    (6, 0)
    (7, 0)
    (8, 0)
    (9, 0.135115958035331)
    (10, 2.5906253109508043)
    (11, 0)
    (12, 6.062441262631683)
};
\addlegendentry{Profit const. \eqref{eq:nonnegative_profits_cons} on the horizon}

\addplot+[
    mark=o,
    color=purple,
    line width=0.6pt,
] coordinates {
    (1,149.2047)
    (2,121.7878)
    (3,101.9810)
    (4,62.3725)
    (5,8.8725)
    (6,0)
    (7,0)
    (8,0)
    (9,86.6281)
    (10,86.1286)
    (11,0)
    (12,16.2658)
};
\addlegendentry{Profit const. \eqref{eq:nonnegative_profits_cons} on each hour}

\end{axis}
\end{tikzpicture}

\vspace{-0.4cm}

\caption{Hourly energy and SCC price profiles (upper) and total
profits of each thermal unit (lower) when profit constraint \eqref{eq:nonnegative_profits_cons} applies to the whole horizon. The SCC prices at the
remaining buses are zero.}
\label{fig:price_profit_over_horizon}
\vspace{-0.3cm}
\end{figure}

The previous study guarantees non-negative profit for generators in each dispatch interval. This subsection replaces these hourly constraints and restricts only the total profit over the full daily scheduling horizon to be non-negative. The obtained market clearing prices and generator profits are presented in Fig.~\ref{fig:price_profit_over_horizon}. Compared to the case without profit constraints (Table~\ref{table:solving_time_gap_multiperiod}), the resulting primal objective rises by only \(0.23\%\), while the dual objective remains unchanged.

Restricting only the total profit over the horizon would eliminate the strong coupling between energy prices and SCC prices observed in Fig.~\ref{fig:price_split}, since prices can adjust more flexibly to remunerate units, thus yielding a smaller P-D gap. For instance, the energy and SCC prices at 03:00 are both lower than those in Fig.~\ref{fig:price_split}, leading to losses ranging from \(-2434\) € to \(-558\) € for all generators except $2g$-$b4$, $2g$-$b5$ and $g_1$-$b30$. Such losses are however permissible under this formulation, decreasing the sum of profits by $41.91\%$.

This analysis suggests that the activation scope of the non-negative profit constraint should be selected according to specific market rules. For example, with a signed contract covering a required period, generating units would follow the dispatch as long as they avoid cumulative losses. However, if service providers require continuous profitability, the system operator would need to slightly adjust the scheduling and market clearing to implement hourly profit constraints.

\subsubsection{Dispatch Efficiency Loss vs. Avoided Make-Whole Payments}

\begin{table}[t]
\centering
\caption{Hourly Profit Shortfalls of Thermal Units \\ when Primal-Dual Dispatch Deviates (\texteuro/h)}
\vspace{-0.3cm}
\setlength{\tabcolsep}{2pt}
{\fontsize{8pt}{12pt}\selectfont
\begin{tabular}{lcccccccccc}
\toprule
Time  & 02:00 & 03:00 & 04:00 & 05:00 & 06:00 & 08:00 & 10:00 & 21:00 & 23:00 & 24:00 \\
\midrule
$g_2$-$b3$  & $\mathrm{325}$ & $\mathrm{290}$ & $\mathrm{277}$ & $\mathrm{312}$ & $\mathrm{332}$ & -- & -- & -- & $\mathrm{374}$ & $\mathrm{392}$ \\
$g_1$-$b4$  & -- & -- & -- & -- & -- & $\mathrm{\num{10083}}$ & -- & $\mathrm{386}$ & -- & -- \\
$g_1$-$b30$ & -- & -- & -- & -- & -- & -- & $\mathrm{329}$ & -- & -- & -- \\
$g_2$-$b30$ & -- & -- & -- & -- & -- & -- & $\mathrm{310}$ & -- & -- & -- \\
\bottomrule
\end{tabular}
}
\label{table:make_whole_payment}
\vspace{-0.5cm}
\end{table}

Note that the market outcomes in Fig.~\ref{fig:profit_12units} are settled based on the dispatch obtained from the P-D formulation (total cost: \(1.308\) M\texteuro~in Table~\ref{table:solving_time_gap_multiperiod}), instead of the cost-minimal schedule of \(1.301\) M\texteuro. If the optimal dispatch is instead used for settlement, some generating units would incur profit shortfalls, as summarized in Table~\ref{table:make_whole_payment}. Unit $g_1$-$b4$ exhibits the largest shortfall at 08:00, reaching $0.01$ M\texteuro, while smaller deficits appear for all other units across several dispatch intervals. However, when cumulative profits over the horizon are considered instead, all units achieve positive final profits.

Given that the P-D pricing formulation permits and would implement suboptimal dispatch, this case further demonstrates that the dispatch efficiency loss introduced by constraint \eqref{eq:nonnegative_profits_cons} can be modest (\(0.54\%\) here). This small rise in system cost is traded for the elimination of the make-whole payments that would otherwise be required to compensate for the aforementioned profit shortfalls, totaling $0.013$ M\texteuro.

\vspace{-0.3cm}
\subsection{Overview of Market Settlement}\label{Overview of Market Settlement}
This subsection briefly analyzes the settlement for energy and SCC ancillary services proposed in this work.

First, consider energy trading. The energy revenue for IBR is `$\lambda^{\mathrm{E}} P_c$', while that for SGs is `$\lambda^{\mathrm{E}} P_g$'. Their combined energy revenue exactly equals the total energy expenditure of all electricity consumers, i.e., `$\lambda^{\mathrm{E}} \mathrm{P}^\mathrm{D}$', as in \eqref{eq:primal_cons_power_balance}.

Regarding SCC trading, the SCC revenue earned by SG $g$ is given in \eqref{eq:nonnegative_profits_cons}, i.e., `$(\sum_b \mathrm{k}_{bg} u_g + \sum_b \sum_{ \{m | g \in m \} } \mathrm{k}_{bm} \eta_m)\lambda_b^\mathrm{SCC}$'. Similarly, the SCC revenue earned by IBR $c$ can be expressed as `$\sum_{b} \lambda_{b}^{\mathrm{SCC}} \mathrm{k}_{bc}\alpha_{c}$'. Both revenue streams aggregate the charges collected from all buses purchasing SCC services. As for the allocation of these costs, let us assume that the loads at these buses bear the costs. Accordingly, the total SCC payment at bus $b$ is given by `$(\sum_{g}\mathrm{k}_{bg}u_{g}+\sum_{m}\mathrm{k}_{bm}\eta_{m}+\sum_{c}\mathrm{k}_{bc}\alpha_{c})\lambda_b^\mathrm{SCC}$'. Under this allocation scheme, the sum of SCC payments across all buses equals the total SCC-related revenues obtained by all units.

However, designing a dedicated SCC cost-allocation mechanism warrants further investigation, since it is a complex issue involving considerations of cost causation, the avoidance of cross-subsidies, and the incentives for market participants. As demonstrated in \cite{matamala2024cost} for inertia and frequency response services, different allocation methodologies can lead to substantially different outcomes. A comprehensive investigation of SCC cost allocation is therefore beyond the scope of this paper and is left for future research.

\vspace{-0.4cm}
\subsection{Sensitivity of Pricing to Binary Expansion}\label{Sensitivity of Pricing to Binary Expansion Errors}
This subsection investigates the impact of pricing outcomes on parameter settings in the binary expansion. For the analysis of the typical cases, a load level of 5.6 GWh is uniformly selected here.

\subsubsection{On Number of Segments}\label{On Number of Segments}

\begin{figure}[!t]
\centering
\begin{tikzpicture}
\begin{axis}[
    width=0.9\linewidth, height=4cm,
    xlabel={\scriptsize Binary expansion parameter $\mathrm{N}$},
    ylabel={\scriptsize Objective values (k\texteuro)},
    label style={font=\scriptsize}, tick label style={font=\scriptsize},
    xmin=8, xmax=32,
    ymin=55.0, ymax=58.0,
    xtick={8,12,16,20,24,28,32},
    ytick={55,56,57,58},
    grid=both, grid style={dotted, gray!60},
    legend style={
        font=\scriptsize, 
        fill=none, 
        draw=none, 
        at={(0.155,0.55)}, anchor=north
    },
    legend cell align={left},
    legend image post style={scale=1},
    legend image code/.code={%
        \draw[mark repeat=1,mark phase=1,#1] plot coordinates {(0,0)};
    },
    axis y line*=left,
    axis x line*=bottom,
    axis lines=box
]
\addplot+[color=green!50!black, mark=o, line width=0.6pt]
    coordinates {
        (8, 57.34221)
        (12,57.33565)
        (16,57.33570)
        (20,57.33570)
        (24,57.33579)
        (28,57.33578)
        (32,57.33565)
    };
\addlegendentry{Primal obj.}

\addplot+[color=blue, mark=square*, line width=0.6pt]
    coordinates {
        (8, 55.27450)
        (12,55.27453)
        (16,55.27451)
        (20,55.27453)
        (24,55.27453)
        (28,55.27452)
        (32,55.27453)
    };
\addlegendentry{Dual obj.}
\end{axis}

\begin{axis}[
    width=0.9\linewidth, height=4cm,
    xmin=8, xmax=32,
    ymin=0.2, ymax=0.5,
    axis y line*=right,
    axis x line=none,
    ylabel={\scriptsize Computation time (s)},
    xtick={8,12,16,20,24,28,32},
    ytick={0.2,0.3,0.4,0.5},
    tick label style={font=\scriptsize},
    tick align=outside,
    ylabel style={at={(axis description cs:1.15,0.5)}},
    legend style={
        font=\scriptsize, 
        fill=none, 
        draw=none, 
        at={(0.105,0.7)}, anchor=north,
        row sep=0pt,
    },
    legend cell align={left},
    legend image post style={scale=1},
    legend image code/.code={%
        \draw[mark repeat=1,mark phase=1,#1] plot coordinates {(0,0)};
    }
]
\addplot+[color=red!70!black, mark=triangle*, line width=0.6pt]
    coordinates {
        (8, 0.40)
        (12,0.41)
        (16,0.45)
        (20,0.39)
        (24,0.34)
        (28,0.27)
        (32,0.41)
    };
\addlegendentry{Time}
\end{axis}
\end{tikzpicture}
\vspace{-0.9cm}
\caption{Primal and dual objective values and computation time versus the binary expansion parameter $\mathrm{N}$ in the P-D pricing formulation.
}
\label{fig:sensitivity_N}
\vspace{-0.3cm}
\end{figure}

The discretization parameter `$\mathrm{N}$' in \eqref{eq:binary_expansion_2} governs a trade‑off between approximation fidelity and computational complexity. For each generator and each time period, the binary expansion introduces $\mathrm{N}$ new binary variables $s_n, (n=1,\dots,\mathrm{N})$ and $8\mathrm{N}$ additional big-M constraints: $4\mathrm{N}$ for linearizing `$\lambda^\mathrm{E} s_n$' and another $4\mathrm{N}$ for `$\lambda^\mathrm{E}u_g s_n$'. It is thus relevant to perform sensitivity analysis over this parameter. For the test system, we examine the variations in the P-D solution and the pricing computation time for $\mathrm{N}=\{8,12,16,20,24,28,32\}$. The results are shown in Fig.~\ref{fig:sensitivity_N}, with MIP optimality gaps of $\{0, 8, 31, 34, 73, 65, 3\} \cdot 10^{-6}$, which are below the solver tolerance of \(10^{-4}\).

The primal and dual objective values remain nearly constant as $\mathrm{N}$ increases, demonstrating that the approximation error introduced by binary expansion would exert no substantial impact on the resulting pricing outcomes. Meanwhile, solver time does not exhibit an upward trend with larger $\mathrm{N}$, indicating that the choice of $\mathrm{N}$ would not be the primary source of computational burden for binary expansion within the investigated range. Therefore, a moderate value (like `$\mathrm{20}$' chosen in this paper) of $\mathrm{N}$ suffices to deliver satisfactory accuracy for the proposed P-D pricing framework.

\subsubsection{On Upper Bounds of Prices}

\begin{figure}[t]
\centering
\begin{tikzpicture}

\begin{axis}[
    width=0.5\linewidth, height=4cm,
    at={(-0.08\linewidth,0)}, anchor=south west,
    xlabel={\scriptsize Energy price cap scaling factor},
    ylabel={\scriptsize Energy price (\texteuro/MWh)},
    label style={font=\scriptsize},
    tick label style={font=\scriptsize},
    xmin=1, xmax=3,
    ymin=8.0, ymax=8.9,
    xtick={1.0,1.6,2.2,2.8},
    ytick={8.0,8.2,8.4,8.6,8.8},
    grid=both, grid style={dotted, gray!60},
    legend style={
        font=\scriptsize,
        fill=none,
        draw=none,
        at={(0.51,0.95)}, anchor=north
    },
    legend cell align={left},
    legend image post style={scale=1},
    legend image code/.code={%
        \draw[mark repeat=1,mark phase=1,#1]
        plot coordinates {(0,0)};
    },
    axis y line*=left,
    axis x line*=bottom,
    axis lines=box,
        ylabel style={at={(axis description cs:-0.15,0.5)}}
]

\addplot+[color=green!50!black, mark=o, line width=0.6pt]
coordinates {
    (1.0,8.72305)
    (1.2,8.72000)
    (1.4,8.80720)
    (1.6,8.33548)
    (1.8,8.10118)
    (2.0,8.10120)
    (2.2,8.10120)
    (2.4,8.10126)
    (2.6,8.10121)
    (2.8,8.10121)
    (3.0,8.10120)
};
\addlegendentry{$\lambda^\mathrm{E}$}
\end{axis}

\begin{axis}[
    width=0.5\linewidth, height=4cm,
    at={(-0.08\linewidth,0)}, anchor=south west,
    xmin=1, xmax=3,
    ymin=1.8, ymax=4.8,
    axis y line*=right,
    axis x line=none,
    ylabel={\scriptsize P-D gap (k\texteuro)},
    xtick={1.0,1.6,2.2,2.8},
    ytick={2,2.5,3,3.5,4,4.5},
    tick label style={font=\scriptsize},
    tick align=outside,
    ylabel style={at={(axis description cs:1.2,0.5)}},
    legend style={
        font=\scriptsize,
        fill=none,
        draw=none,
        at={(0.83,0.8)}, anchor=north east,
        row sep=0pt
    },
    legend cell align={left},
    legend image post style={scale=1},
    legend image code/.code={%
        \draw[mark repeat=1,mark phase=1,#1]
        plot coordinates {(0,0)};
    }
]

\addplot+[color=blue, mark=square*, line width=0.6pt]
coordinates {
    (1.0,4.54989)
    (1.2,3.55850)
    (1.4,2.67201)
    (1.6,2.17784)
    (1.8,2.06127)
    (2.0,2.06117)
    (2.2,2.06112)
    (2.4,2.06126)
    (2.6,2.06122)
    (2.8,2.06116)
    (3.0,2.06112)
};
\addlegendentry{P-D gap}
\end{axis}

\begin{axis}[
    width=0.5\linewidth, height=4cm,
    at={(0.47\linewidth,0)}, anchor=south west,
    xlabel={\scriptsize Energy price cap scaling factor},
    ylabel={\scriptsize SCC price (k\texteuro/p.u.)},
    label style={font=\scriptsize},
    tick label style={font=\scriptsize},
    xmin=1, xmax=3,
    ymin=0, ymax=1.8,
    xtick={1.0,1.6,2.2,2.8},
    ytick={0,0.3,0.6,0.9,1.2,1.5,1.8},
    grid=both, grid style={dotted, gray!60},
    legend style={
        font=\scriptsize,
        fill=none,
        draw=none,
        at={(0.74,0.85)}, anchor=north
    },
    legend cell align={left},
    legend image post style={scale=1},
    legend image code/.code={%
        \draw[mark repeat=1,mark phase=1,#1]
        plot coordinates {(0,0)};
    },
    axis lines=box,
        ylabel style={at={(axis description cs:-0.15,0.5)}}
]

\addplot+[color=green!50!black, mark=o, line width=0.6pt]
coordinates {
    (1.0,0.869724)
    (1.2,1.043668)
    (1.4,1.217612)
    (1.6,1.391560)
    (1.8,1.511588)
    (2.0,1.511588)
    (2.2,1.511588)
    (2.4,1.511584)
    (2.6,1.511588)
    (2.8,1.511588)
    (3.0,1.511588)
};
\addlegendentry{$\lambda^\mathrm{SCC}_{26}~(/25)$}

\addplot+[color=blue, mark=square*, line width=0.6pt]
coordinates {
    (1.0,0.839923)
    (1.2,0.775469)
    (1.4,0.520378)
    (1.6,0.0)
    (1.8,0.0)
    (2.0,0.0)
    (2.2,0.0)
    (2.4,0.0)
    (2.6,0.0)
    (2.8,0.0)
    (3.0,0.0)
};
\addlegendentry{$\lambda^\mathrm{SCC}_{29}$}

\addplot+[color=red!70!black, mark=triangle*, line width=0.6pt]
coordinates {
    (1.0,0.289879)
    (1.2,0.0627495)
    (1.4,0)
    (1.6,0.188222)
    (1.8,0.0)
    (2.0,0.0)
    (2.2,0.0)
    (2.4,0.0)
    (2.6,0.0)
    (2.8,0.0)
    (3.0,0.0)
};
\addlegendentry{$\lambda^\mathrm{SCC}_{30}$}

\end{axis}

\end{tikzpicture}

\vspace{-0.45cm}
\caption{Variations of energy price and P-D gap (left), and SCC prices (right) with different energy price cap scaling factors. All other SCC prices are zero.}
\label{fig:energy_price_gap_scc}
\vspace{-0.4cm}
\end{figure}

As stated in Section~\ref{Primal-Dual Formulation}, the uniform price caps used in the previous case studies are set to twice the relevant unit costs, i.e., `$2 \cdot \mathrm{max}\{\mathrm{c}^\mathrm{m}_g~|~\forall g \}$' for energy and `$2 \cdot \mathrm{max} \{ \mathrm{c}_g^\mathrm{nl} + \mathrm{c}_g^\mathrm{st}~|~\forall g  \}$' for SCC. The sensitivity analysis here investigates the impact of this scaling coefficient, which varies from 1 to 3 in increments of 0.2. The results obtained by adjusting the energy price cap are presented in Fig.~\ref{fig:energy_price_gap_scc}. Corresponding results for the SCC cases are not plotted, since varying the SCC price cap produces no observable difference, demonstrating that the computed SCC prices never reach the imposed upper bound.

As shown in Fig.~\ref{fig:energy_price_gap_scc}, both the energy price and the P-D gap (left plot) gradually decrease as the energy price cap increases and remain constant once the scaling factor reaches two. This indicates that the price cap constraint becomes non-binding beyond this threshold, yielding stable market clearing outcomes. This behavior also applies to SCC cases (right plot). Specifically, for small scaling factors, energy prices would be defined inappropriately, which would distort SCC price signals (redundant signals for buses 29 and 30) and lead to suboptimal scheduling and degraded market clearing efficiency.

Based on the above analysis, the scaling factor of `$2$' is selected, as it defines an upper bound that is never reached by the prices while maintaining a reasonably sized feasible region for the problem solution.

\vspace{-0.4cm}
\subsection{Applicability of the Primal-Dual Pricing for SCC Services}\label{Applicability of the Primal-Dual Pricing for SCC Services}
This subsection discusses the impact of approximation errors in SCC constraints on post-pricing market outcomes and evaluates the scalability of the proposed pricing framework.

\subsubsection{On Errors of the Approximation}
The Mean Percentage Error (MPE) of SCC constraint coefficients for the IEEE 30-bus system, evaluated across all operating conditions, is $-1.79\%$. Computed over samples satisfying `$I_{b_\mathrm{L}} < \mathrm{I}_{b_{\mathrm{lim}}} \cap I_{b_\mathrm{SC}} \ge \mathrm{I}_{b_{\mathrm{lim}}}$', this error shows that the approximated SCC provides a tighter security boundary than the grid-code limit, thereby ensuring system security. This error level is permissible and acceptable and verifies the effectiveness of the approximation \cite{wang2026imperfect}.

For the test cases in Sections~\ref{Single-Period SCC Service Pricing} and \ref{Multi-Period SCC Service Pricing}, the approximation error is zero because, given the commitment decisions, both approximated $I_{b_\mathrm{L}}$ and exact $I_{b_\mathrm{SC}}$ are enforced above $I_{b_\mathrm{lim}}$, and $I_{b_\mathrm{SC}}$ is typically closer to the limit. In the single-period case, the maximum MPE between approximated and exact SCC is $29.28\%$ for bus 29, whereas bus 26 achieves exact matching. For the multi-period case, the maximum MPE is $24.31\%$ at bus 29 as well, and the minimum MPE of $0.02\%$ occurs at bus 26. Importantly, the SCC shadow price `$\lambda^\mathrm{SCC}_{26}$' is only active (as in Figs.~\ref{fig:price_scc_demand} and \ref{fig:price_split}), where the approximation error is negligible. Therefore, these deviations would not raise market concerns, as only bus 26 purchases SCC services in these cases. As a result, there would be no erroneous ancillary service charges, and SGs would neither overearn nor underearn revenue caused by the approximation discrepancy.

\subsubsection{On Scalability of Primal-Dual Pricing}

\begin{table}[t]
\centering
\caption{Computation Time for Approximating Short-Circuit Current Constraints and Single-Period Pricing (seconds)}
\vspace{-0.2cm}
\setlength{\tabcolsep}{3.5pt}
{\fontsize{8pt}{12pt}\selectfont
\begin{tabular}{lccc}
\toprule
\qquad \quad Test & \begin{tabular}[c]{@{}c@{}}Dataset\\ generation\end{tabular} & \begin{tabular}[c]{@{}c@{}}Offline\\ training\end{tabular} & \begin{tabular}[c]{@{}c@{}}P-D\\ pricing\end{tabular} \\
\midrule
Modified 30-bus system (12 SGs)  & $\mathrm{19}$ & $\mathrm{105}$ & $\mathrm{0.41}$ \\
Modified 57-bus system (14 SGs) & $\mathrm{441}$ & $\mathrm{2053}$  & $\mathrm{0.53}$ \\
Modified 118-bus system (12 SGs) & $\mathrm{1875}$ & $\mathrm{545}$ & $\mathrm{0.47}$ \\
\bottomrule
\end{tabular}
}
\label{table:scalability_time}
\vspace{-0.4cm}
\end{table}

As described in Section~\ref{Formulation of Short Circuit Current Constraints}, training the SCC constraints by enumerating all unit operating states generates \(2^{|\mathcal{G}|+|\mathcal{C}|}\) samples in set \(\Omega\) (the IEEE 30-bus system here requires \(2^{15}\) samples), while the pairwise terms in set \(\mathcal{M}\) grow as the combinatorial number \(C_{|\mathcal{G}|}^{2}\). As the number of units increases, the sample size grows exponentially and the number of pairwise terms grows quadratically. 

To evaluate scalability, the P-D pricing is additionally applied to modified IEEE 57-bus and 118-bus systems, retaining the wind conditions and generator characteristics from the previous cases. The computational performance for SCC constraint approximation and pricing is summarized in Table~\ref{table:scalability_time}. As the system size increases, dataset generation becomes more time-consuming due to the inverse operation for larger admittance matrices. Offline training also takes longer, particularly for the 57-bus system, which involves more SGs and thus a larger optimization problem and more classification conditions. Nevertheless, the computational burden remains manageable, as SCC constraints need to be approximated only once per system and training can be accelerated using powerful computing resources. By contrast, pricing computation is less sensitive to system size and would remain relatively stable within the considered range of SG numbers, which is the main driver for the number of binary variables needed in the P-D formulation.

\section{Conclusion and Future Work}\label{Conclusion}
Given the limitations of existing SCC pricing models in handling binary variables, such as inadequate remuneration and requirement for make-whole payments, a primal-dual formulation has been adopted, for the first time, to offer new insights on how to effectively compute the shadow price of SCC services while preserving the UC nature. Compared with the dispatchable method, this approach
avoids spurious price signals at SCC-irrelevant buses, and produces sufficient prices that eliminate make-whole payments. The restricted method is simply not suitable for remunerating SCC provided by synchronous condensers, since they usually offer multiple discrete services, which, however, cannot be separately priced. Even for thermal units, whose SCC contribution may be priced using this method, it has been shown that it would lead to unintuitive SCC prices due to its coupling with the commitment decisions, and uplift payments may still be required.

In short, to retain non-convexities and obtain interpretable prices for the SCC service, the primal-dual formulation is demonstrated to be an effective way to achieve these goals. In future work, a holistic pricing framework which includes other ancillary services involving binary variables should be developed, as it may be non-trivial to extend the primal-dual formulation to other services, such as voltage stability.

{\appendices
    
\section{Dual Formulation of McCormick Envelopes}\label{Dual constraints for UC states in McCormick envelopes}
This section derives the dual term $h_g(\gamma_{m,1}^{\mathrm{max}}, \gamma_{m,2}^{\mathrm{max}}, \gamma_{m,1}^{\mathrm{min}})$ in \eqref{eq:dual_cons_binary_1}. Taking $g_1\text{-}b2$ and $g_2\text{-}b2$ as an example, the composition of $h_{g_1\text{-}b2}(\gamma_{m,1}^{\mathrm{max}}, \gamma_{m,2}^{\mathrm{max}}, \gamma_{m,1}^{\mathrm{min}})$ and $h_{g_2\text{-}b2}(\gamma_{m,1}^{\mathrm{max}}, \gamma_{m,2}^{\mathrm{max}}, \gamma_{m,1}^{\mathrm{min}})$ is illustrated as follows:
\begin{subequations} \label{eq:h_g}
\begin{align}
& h_{g_1\text{-}b2}(\gamma_{m,1}^{\mathrm{max}}, \gamma_{m,2}^{\mathrm{max}}, \gamma_{m,1}^{\mathrm{min}}) 
= - \sum_{m=1}^{11}\gamma^{\mathrm{max}}_{m,1} + \sum_{m=1}^{11}\gamma^{\mathrm{min}}_{m,1}   \label{eq:eq:h_g12} \\
& h_{g_2\text{-}b2}(\gamma_{m,1}^{\mathrm{max}}, \gamma_{m,2}^{\mathrm{max}}, \gamma_{m,1}^{\mathrm{min}}) 
= - \gamma^{\mathrm{max}}_{1,2} + \gamma^{\mathrm{min}}_{1,1} \nonumber \\
& \hspace{3.8cm} - \sum_{m=12}^{21}\gamma^{\mathrm{max}}_{m,1} + \sum_{m=12}^{21}\gamma^{\mathrm{min}}_{m,1} \label{eq:h_g22} 
\end{align}
\end{subequations}

The cases for other generators' relaxed commitment variables can be written in the same manner as above. For each possible combination of two commitment variables in set $\mathcal{M}$, four auxiliary constraints (\eqref{eq:MC_linear_1}–\eqref{eq:MC_linear_3} and \eqref{eq:MC_linear_5}) with index $m$ must be added to the primal problem, resulting in a total number of combinations of $\lvert \mathcal{M} \rvert = C_{|\mathcal{G}|}^{2} = 66$. The reader is referred to \cite{Code} for details.

\vspace{-0.3cm}
\section{Multi-Period Primal-Dual Formulation}\label{Multi-Period Primal-Dual Formulation for Pricing Short-Circuit Current Services}
Extending the single-period P-D formulation to a multi-period setting primarily restates constraints that couple commitment variables `$u_{g,t}$' across consecutive time steps, such as the startup constraint and its associated dual term. After introducing a time index `$t$' into the whole model, i.e., \eqref{eq:nonnegative_profits_cons} and \eqref{eq:final_pricing}, the key expressions that need to be restated are presented in \eqref{eq:multiperiod_PD}, while all other components remain unchanged.
\begin{subequations}  \label{eq:multiperiod_PD}
\begin{align}
 & \eqref{eq:primal_cons_st_cost_lb} \Rightarrow C_{g,t}^\mathrm{st} \ge (u_{g,t}-u_{g,t-1})\mathrm{c}^\mathrm{st}_{g}:  (\sigma_{g,t}^\mathrm{st}),~ \forall g,t \label{eq:primal_cons_st_cost_lb_multiperiod} \\
 &  \eqref{eq:dual_obj} \Rightarrow \displaystyle \max_{V_\mathrm{D}}  ~ \sum_t \Big( \mathrm{P}_t^\mathrm{D}\lambda^{\mathrm{E}}_t+\sum_b(\mathrm{I}_{b_{\mathrm{lim}}}-\sum_{c}\mathrm{k}_{bc}\alpha_{c,t})\lambda_{b,t}^{\mathrm{SCC}} \nonumber \\
     & \hspace{1.2cm} -\sum_c\alpha_{c,t}\mathrm{P}^{\mathrm{max}}_{c,t}\zeta_{c,t}^\mathrm{max} -  \sum_{g}\psi_{g,t}^{\mathrm{max}} -\sum_{m}\gamma_{m,1,t}^{\mathrm{min}} \Big) \nonumber \\
     & \hspace{1.2cm} - \sum_{g}  u_{g,0}\mathrm{c}_{g}^{\mathrm{st}}\sigma_{g,t=1}^{\mathrm{st}} \label{eq:dual_0bj_multiperiod} \\
& \eqref{eq:dual_cons_binary_1} \Rightarrow \mathrm{c}_{g}^\mathrm{nl} -\sum_b\mathrm{k}_{bg}\lambda_{b,t}^{\mathrm{SCC}}-\mathrm{P}_{g}^{\mathrm{max}}\mu_{g,t}^{\mathrm{max}}+\mathrm{P}_{g}^{\mathrm{min}}\mu_{g,t}^{\mathrm{min}} \nonumber \\
    & \hspace{1.2cm} + \mathrm{c}_{g}^{\mathrm{st}}\sigma_{g,t}^{\mathrm{st}} 
    - \mathrm{c}_{g}^{\mathrm{st}}\sigma_{g,t+1}^{\mathrm{st}} +h_g(\gamma_{m,1,t}^{\mathrm{max}}, \gamma_{m,2,t}^{\mathrm{max}}, \gamma_{m,1,t}^{\mathrm{min}}) \nonumber \\
    & \hspace{1.2cm} +\psi_{g,t}^{\mathrm{max}} \ge 0,~\forall g, t \leq T-1 \label{eq:dual_cons_commit)multiperiod}
\end{align}
\end{subequations}
where the term `$\mathrm{c}_{g}^{\mathrm{st}}\sigma_{g,t+1}^{\mathrm{st}}$' in \eqref{eq:dual_cons_commit)multiperiod} will be zero for `$t = T$', i.e., the last period over the market horizon.

Furthermore, linear constraints including ramp-rate and minimum up/down time limits can also be embedded within the MILP \eqref{eq:final_pricing} to achieve enhanced modeling. Adding these constraints introduces new dual terms in the objective \eqref{eq:dual_0bj_multiperiod}. Dual variables from ramp limits enter \eqref{eq:dual_cons_Pg_2} in a multi-period manner, while those for minimum up/down time constraints appear in \eqref{eq:dual_cons_commit)multiperiod} and produce an additional set of constraints. Further details can be found in \cite{ye2019incorporating}.

}

\vspace{-0.3cm}
\bibliographystyle{IEEEtran}
\bibliography{main}

\end{document}